\documentclass[final,5p,times, twocolumn]{elsarticle}
\usepackage{graphicx}
\usepackage{amssymb}
\usepackage[dvips]{psfrag}
\usepackage[latin1]{inputenc}
\usepackage{url}
\usepackage{eurosym}

\usepackage[nodots]{numcompress}

\usepackage{endnotes}

\usepackage{threeparttable}
\usepackage{color}

\journal{Renewable Energy}

\begin{document}

\begin{frontmatter}

%% Title, authors and addresses

%% use the tnoteref command within \title for footnotes;
%% use the tnotetext command for the associated footnote;
%% use the fnref command within \author or \address for footnotes;
%% use the fntext command for the associated footnote;
%% use the corref command within \author for corresponding author footnotes;
%% use the cortext command for the associated footnote;
%% use the ead command for the email address,
%% and the form \ead[url] for the home page:
%%
%% \title{Title\tnoteref{label1}}
%% \tnotetext[label1]{}
%% \author{Name\corref{cor1}\fnref{label2}}
%% \ead{email address}
%% \ead[url]{home page}
%% \fntext[label2]{}
%% \cortext[cor1]{}
%% \address{Address\fnref{label3}}
%% \fntext[label3]{}

\title{The Global Grid}
%\title{DRAFT}
%\title{The Global Grid: A New Concept}
%\title{The Global Grid: New Opportunities and Challenges}

%% use optional labels to link authors explicitly to addresses:
%% \author[label1,label2]{<author name>}
%% \address[label1]{<address>}
%% \address[label2]{<address>}

%\author[ETH]{Spyros Chatzivasileiadis}
%\ead{spyros@eeh.ee.ethz.ch}
%\author[Ulg]{Damien Ernst}
%\ead{dernst@ulg.ac.be}
%\author[ETH]{G\"{o}ran Andersson}
%\ead{andersson@eeh.ee.ethz.ch}

\author[ETH]{Spyros Chatzivasileiadis\corref{cor1}\fnref{eehwebsite}}
\ead{spyros@eeh.ee.ethz.ch}
\author[Ulg]{Damien Ernst\fnref{damientel}}
\ead{dernst@ulg.ac.be}
\author[ETH]{G\"{o}ran Andersson\fnref{eehwebsite}}
\ead{andersson@eeh.ee.ethz.ch}

\cortext[cor1]{Corresponding author. Tel.:+41 446328990. Fax:+41 446321252.}
\fntext[damientel]{Tel:+32 43669518; www.montefiore.ulg.ac.be/$\sim$ernst/}
\fntext[eehwebsite]{www.eeh.ee.ethz.ch}
\address[ETH]{Power Systems Laboratory, ETH Zurich, 8092 Zurich, Switzerland}
\address[Ulg]{Institut Montefiore, University of Li\`ege, 4000 Li\`ege, Belgium}

\begin{abstract}
This paper puts forward the vision that a natural future stage of the electricity network could be a grid spanning the whole planet and connecting most of the large power plants in the world: this is the ``Global Grid''. The main driving force behind the Global Grid will be the harvesting of remote renewable sources, and its key infrastructure element will be the high capacity long transmission lines. Wind farms and solar power plants will supply load centers with green power over long distances.

This paper focuses on the introduction of the concept, showing that a globally interconnected network can be technologically feasible and economically competitive. We further highlight the multiple opportunities emerging from a global electricity network such as smoothing the renewable energy supply and electricity demand, reducing the need for bulk storage, and reducing the volatility of the energy prices. We also discuss possible investment mechanisms and operating schemes. Among others, we envision in such a system a global power market and the establishment of two new coordinating bodies, the ``Global Regulator'' and the ``Global System Operator''.

\end{abstract}

\begin{keyword}
%% keywords here, in the form: keyword \sep keyword
global electricity grid \sep electricity transmission network \sep global electrical network \sep renewable energy \sep wind power \sep solar power
%% MSC codes here, in the form: \MSC code \sep code
%% or \MSC[2008] code \sep code (2000 is the default)
\end{keyword}

\end{frontmatter}

\section{Introduction}

Increased environmental awareness has led to concrete actions in the energy sector in recent years. Examples are the European Commission's target of 20\% participation of renewable energy sources (RES) in the EU energy mix by 2020 \cite{EC_202020} and California's decision to increase renewable energy in the state's electricity mix to 33\% of retail sales, again by 2020 \cite{California_energybill}. At the same time, several studies have been carried out investigating the possibilities of a higher share of renewables in the energy supply system of the future. For instance, the German Energy Agency (DENA) assumes 39\% RES participation by 2020 \cite{dena_study}, while a detailed study from the National Renewable Energy Laboratory suggests that meeting the US electricity demand in 2050 with 80\% RES supply is a feasible option~\cite{NREL_RE_Futures}.
In \cite{Czisch_eurosupergrid} and \cite{Czisch_Book_IET}, a 100\% renewable energy supply system in Europe with interconnections in North Africa and West Asia is discussed. A similar study on a global scale was carried out by WWF and Ecofys in~\cite{WWF_energyreport}. The study concluded that a 100\% renewable energy supply by 2050, although an ambitious goal, is both cost-effective and technically feasible. More recently, \citet{WWSglobal_generation} investigated ``the feasibility of providing worldwide energy for all purposes (electric power, transportation, heating/cooling,etc.) from wind, water, and sunlight''. The authors made a detailed analysis and proposed a plan for implementation. They found that the barriers to the deployment of this plan are not technological or economic, but rather social and political. %\cite{WWSglobal_generation} also mention several other studies which examine 100\% energy supply from renewables in Europe, the %U.S. or Australia.

All these studies suggest that for an efficient integration of more renewable sources in the current system, a reinforcement of the transmission system is necessary in order to reliably satisfy the energy demand. In \cite{dena_study}, the need for constructing 1700-3600~km additional transmission lines in Germany and the neighboring regions is emphasized, in order to avoid non-transmissible power from a 39\% RES penetration in the German electricity system. Towards the same end, the ``Tres Amigas'' project has been initiated in the US in order to interconnect the three US transmission systems and facilitate increased RES integration $\langle$www.tresamigasllc.com$\rangle$. Benefits from interconnection are also pointed out in Ref.~\cite{archer_jacobson}. The authors studied the interconnection of 19 dispersed wind generation sites and found that, on average, 33\% of the yearly averaged wind power can be used with the same reliability as a conventional power plant.

At the same time, long transmission lines are being considered for harvesting renewable energy from remote locations and delivering it to major load centers. Paris et al. seem to have presented the first feasibility analysis of this kind \cite{paris_longtrans}. Later, a study about the construction of a large hydro power plant at the Congo River (Inga Dam) in Central Africa and the transmission of the produced power to Italy was also reported \cite{paris_grandinga}. The conclusion was that such a solution was both feasible and economically competitive. Similarly, in \cite{Iceland_UK}, the profitability of producing electricity from geothermal and hydro power plants in Iceland in order to transmit it and sell it to the UK was demonstrated. Currently, almost 20~years later, the two governments are discussing ways to realize this project~\cite{iceland_uk_guardian}. In \cite{Czisch_Book_IET}, it was also suggested that interconnecting Europe to power plants in regions with higher RES potential such as North Africa, Russia, and West Asia\footnote{The author has estimated the wind power potential of countries such as Kazakhstan, Russia, Mauritania and Morocco in the range of hundreds of Gigawatts for each country (1 Gigawatt = $10^9$ Watt).} could have a cost comparable to the current system. Reference~\cite{Rustec} focused on the Russian RES potential and argued that ``an EU--Russian cooperation in the renewable energy field would present a win-win situation''. Russian renewable energy could help achieve the EU environmental targets, while, at the same time, ``Russia could begin to develop a national renewable energy industry without risking potential price increases for domestic consumers''. Significant network reinforcements, in the form of a Russian--EU Supergrid, would be necessary in such a case.

Concrete actions have been taken to exploit the benefits of interconnections. EU Guidelines already encourage transmission projects such as the Baltic Ring~\cite{Rustec}. Projects such as Desertec $\langle$\url{www.desertec.org}$\rangle$, Medgrid $\langle$\url{www.medgrid-psm.com}$\rangle$, and OffshoreGrid $\langle$\url{www.offshoregrid.eu}$\rangle$ have been launched, in order to interconnect Mediterranean states with Europe and transfer renewable energy from the African deserts or North Sea to the major load centers. At the same time, initiatives such as Gobitec $\langle$\url{www.gobitec.org}$\rangle$ in Asia and Atlantic Wind Connection $\langle$\url{www.atlanticwindconnection.com}$\rangle$ in the USA aim to interconnect the Asian power grids or transmit off-shore wind energy to the US East Coast.

%Initiatives such as  Desertec $\langle$www.desertec.org$\rangle$ and Transgreen~\cite{transgreen}, and projects such as OffshoreGrid $\langle$www.offshoregrid.eu$\rangle$ and Atlantic Wind Connection $\langle$www.atlanticwindconnection.com$\rangle$ have been launched, in order to transfer renewable energy from the African deserts or North Sea to Europe, and from off-shore mid-Atlantic wind farms to the US East Coast.

However, many of these ideas have still remained regional or interregional in nature, concentrating in Europe and its neighboring regions, North America, or Asia. Comparing the electricity network with networks of similar magnitude, such as the transportation or the telecommunications network, one realizes that several of them have already managed to span the globe. It seems that the only network of similar size which does not form interconnections over the world is the electric power grid.

This paper suggests the next logical step for the electricity network: the Global Grid. The energy needs of the Earth's population will continue to grow \cite{weo2010}. In the search for green electricity, new sites will be exploited, even further from the load centers and the current power grids. A point will be reached, where a RES power plant will be in equal distance from two power systems on different continents. A wind farm in Greenland, for instance, would be a realistic example of such a situation. Our analysis in Section~\ref{sec:greenland} shows that connecting such a wind farm to both Europe and North America is a profitable solution. From there, an interconnected global power grid can start to form.

%Larger interconnections can not only deliver this power, but also enhance grid reliability and security of supply, facilitate electricity trade and increase competition.

Searching the literature for similar concepts, in \cite{intercont_grid_russia} and \cite{intercont_grid_hammons} intercontinental interconnections between Russia and North America or Europe and Africa have been discussed, which would eventually lead to a globally interconnected grid. The authors describe the benefits that would arise from tapping unused renewable potential from remote locations, by supplying the consumers with ``cheap'' energy, while they briefly mention opportunities stemming from the time zone and seasonal diversity. They further qualitatively investigate the feasibility of such interconnections and the benefits to society due to the reduced carbon footprint. Similarly, the GENESIS project is described in \cite{Kuwano_genesis}, projecting a world where electricity will be generated from an abundant number of solar photovoltaics, and global interconnections will transmit the power to regions where there is night. Assuming a global electricity grid, the authors in \cite{aboumahboub_Global_EU_grid} model the global solar and wind patterns based on realistic data and present simulation results. Their focus is on the optimal generation mix for a 100\% sustainable electricity supply. Along with these studies, an initiative launched by the Global Energy Network Institute was developed to support such concepts, demonstrating the benefits of global interconnections and compiling related material $\langle$\url{www.geni.org}$\rangle$.

This paper introduces the concept from a more technological point of view, focusing on the transmission grid, while it also refers to transmission investments and network operation schemes. After an illustration of the concept as we envision it in Section~\ref{sec:illustration}, Section~\ref{sec:opportunities} will highlight the emerging opportunities for the Global Grid from a power engineering perspective. Section~\ref{sec:investments} will discuss transmission investments in a global grid environment, while Section~\ref{sec:operation} speculates about possible operational schemes based on the new market structures. After a brief discussion in Section~\ref{sec:discussion}, the paper concludes with Section~\ref{sec:conclusions}.

\section{The Global Grid: An Illustration}
\begin{figure*}[htb]
    \centering
    \includegraphics[width=1\textwidth]{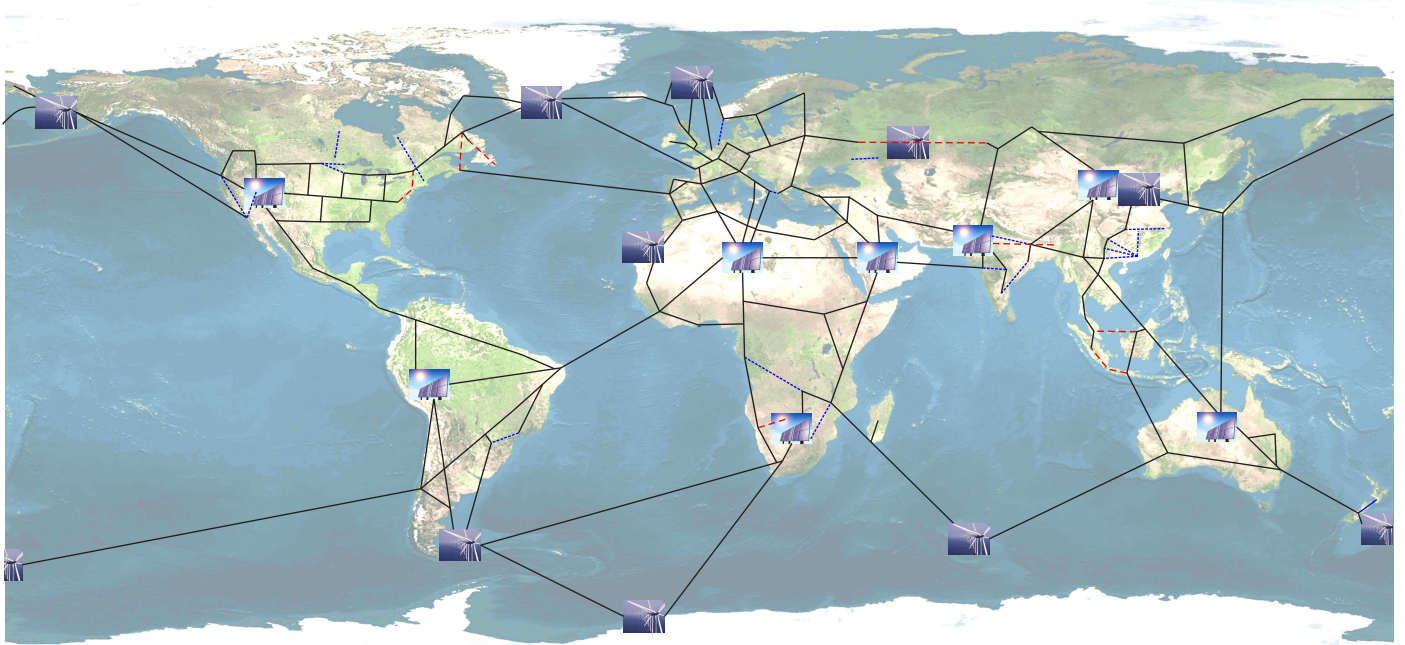}
    \caption{Illustration of a possible Global Grid. The blue dotted lines indicate the HVDC lines with a length over 500 km, that are already in operation. The HVDC lines over 500~km currently in the building/planning phase are indicated in dashed red lines (the list with the illustrated HVDC lines is not exhaustive). The location of the RES power plants has been based on solar radiation maps, average wind speeds, and sea depths (see~\ref{sec:maps}).}\label{fig:GlobalGrid_withExisting}
\end{figure*}

%\section{An Illustration}
\label{sec:illustration}
Before continuing our analysis, the current section is devoted to a brief description of the Global Grid as we envision it. This will hopefully produce a better understanding of the proposed concept. Towards this end, a realistic example leading to intercontinental interconnections is also included. Fig.~\ref{fig:GlobalGrid_withExisting} illustrates a possible global grid. Issues pertaining to the power generation and transmission of the global grid are described below.

\subsection{Generation -- Extreme RES}
We envision that the power supply of the Global Grid will depend on renewable energy sources. Large wind potential exists in off-shore locations, while deserts offer great opportunities for solar power plants. With the Global Grid, it will be feasible to connect these remote locations to end consumers. In a term equivalent to ``Extreme Oil''\footnote{Extreme Oil is often used to describe the extraction of oil through unconventional oil fields or processes (e.g., deep oil drilling or oil extraction from the tar sands).} we will refer to such remote renewable power plants as ``Extreme RES.'' We define Extreme RES as RES power plants in locations where the installation is more difficult than in current projects, or RES power plants for which the technology is not yet mature. Examples are airborne wind turbines \cite{airborne_wind}, installations in deep oceans (e.g., Hywind \cite{hywind}) and in Antarctica (e.g.,~\cite{Cathcart_Antarctica}).

\subsection{Transmission}
\label{sec:transmission}
We anticipate that a power supergrid will be built, serving as a transmission backbone. The new grid should be of a meshed nature, interconnecting all regional power systems into one. Interested parties would be able to ``enter'' the Global Grid\footnote{For the sake of readability, we will often abuse the term ``Global Grid'' within this paper to refer only to the new interconnections, necessary for the realization of the concept. However, the ``Global Grid'' vision includes also the currently existing electricity networks, which intends to unify in a global electricity network of the future.} and transmit power almost anywhere in the world.

Most of the lines forming this large network are expected to be High Voltage Direct Current (HVDC) lines or cables. There are three main reasons for this. First of all, HVDC cables are currently the sole solution for submarine long distance transmission. AC cables can be laid for a maximum distance of ca.~60~km without reactive compensation, while current technologies for Gas-Insulated Lines\footnote{Gas-Insulated Lines (GIL) are a means of bulk electric power transmission at extra high voltage. AC conductors are encased in a metallic tube filled with SF$_6$ and N$_2$ gases. They are considered as a complementary solution to AC lines and cables, and are used when conventional AC transmission is not possible, such as densely populated areas, or along longer underground routes \cite{GIL_maxlength}.} do not allow for a distance exceeding 100~km \cite{GIL_maxlength}.  An additional reason is that long-distance HVDC lines have lower thermal losses than AC lines.

The third reason stems from the fact that non-synchronous areas are going to be connected. Power systems in different regions operate at different voltage frequencies. Coupling the systems at an AC level would not only mean the adoption of a global nominal frequency, but also that every regional power system would probably be more susceptible to the failure of a neighboring system. HVDC links can locally handle dynamic security problems, see Section \ref{sec:security}, and act as a firewall for disturbances between the interconnected grids. Nevertheless, on land and within synchronous systems -- or systems wishing to be synchronized -- Ultra-High Voltage AC lines (UHVAC) as well as short segments of Gas-Insulated Lines could also be anticipated (e.g.,~\cite{ucte_europe_russia}).

Several of the technologies required for the Global Grid are already relatively mature. For the rest, development is needed but no significant barriers are anticipated from the technical point of view. For example, additional experience might be necessary for the multi-terminal technology, which can connect several HVDC lines to one node. The development of deep undersea cables for bulk power transmission will also be required. In \cite{eon_uk}, additional technological challenges are identified which also apply to the Global Grid, such as the further development of HVDC circuit breakers, protection and control systems, as well as a standardised operating voltage level. Concerning the operating voltage level, we expect that a set of two or three different voltage levels would be required in a global grid environment.

\subsection{Wind Farm in Greenland}
\label{sec:greenland}
Tapping the renewable potential in Greenland, as mentioned in the Introduction, would be a realistic example of how we could progress to global interconnections. Undoubtedly, global grid interconnections would be constructed at several different locations, and perhaps earlier than a potential route through Greenland. Nevertheless, Greenland was selected here as a representative example due to its renewable potential, its proximity to Iceland, and the fact that it lies at equal distance from both Europe and North America. It should also be noted that all interconnecting sections along this route have comparable lengths or sea depth to currently existing projects (see Appendix B for more details).

Greenland's hydropower gross potential is approximately 800'000~GWh/year \cite{greenland_hydro}, orders of magnitude higher than the country's net electricity consumption of about 300~GWh in 2010 \cite{eia_greenland}. At the same time, the shores of Greenland experience high winds (above 8.0 m/s), while the relatively shallow waters should allow for the installation of off-shore wind farms\footnote{See Figs.~\ref{fig:map_windpotential}~and~\ref{fig:map_elevation}. In Fig.~\ref{fig:map_elevation}, the shores of Greenland are shown to have a depth between 0 and 500~m. We assume that there must be sites near the coast where the water depth will be suitable for a wind farm installation.}. Hydro-power plants exhibit certain advantages over wind farms, as they are considered non-intermittent, with low operating costs and energy storage potential (i.e., pump-hydros). In comparison, for off-shore wind farms a capacity factor of about 40\% should be assumed (as in, e.g., \cite{WWSglobal_grid}). Nevertheless, for this analysis we focus on the wind farm scenario, so as to account for the less favorable case. Thus, we assume that a 3~GW wind farm off the east shores of Greenland is feasible. We further assume that some investors have decided to connect the wind farm with a 3~GW line to Europe through Iceland and the Faroe Islands. It should be noted here that Ref.~\cite{Iceland_UK} has already shown that the Iceland-UK interconnection is a viable option, while the two governments are currently discussing its possible realization~\cite{iceland_uk_guardian}. The question in this analysis is whether a connection with North America would be profitable, taking into account that the wind farm will be able to sell its produced power always at peak price (50\% of the day to Europe and 50\% of the day to North America). Searching for similar ideas, we found that such a concept has already been proposed in \cite{claverton_greenland}, where the author suggests that Greenland's significant wind potential can be tapped and sold to both continents at peak prices. In this paper we carry out a cost-benefit analysis to support this argument. Through our analysis in \ref{sec:transcosts_greenland}, it seems that the costs per delivered kWh would increase by 21\%-25\% in case the wind farm is connected to both continents. If off-peak prices are half of peak prices, the revenues will increase by 31\%-33\%. Meanwhile, the transmission path North UK-Canada will have a transmission capability of about 20~TWh per year, whereas the wind farm only produces about 10~TWh. Opportunities for electricity trade between the continents will emerge, generating a significant amount of additional revenues. With these considerations, connecting the wind farm with both continents could be a profitable investment\footnote{In the following sections, we will often use as an example the connection of Europe to North America, either through a wind farm or by a direct line. The proposed concept, nevertheless, applies to all possible interconnections around the globe, even if not explicitly discussed in this paper.}. Similar results, if not more favorable, are expected for the case of connecting Greenland's potential hydropower plants.

\section{Opportunities}
\label{sec:opportunities}
The main driving force behind the Global Grid will be the harvesting of remote renewable sources. However, going global allows multiple new opportunities to emerge, providing a significant incentive for the successful implementation of this concept. A brief analysis of some of them follows.

\subsection{Smoothing Out Electricity Supply and Demand}
\label{sec:smoothing}
A typical daily load curve has one or two peaks of power demand around midday, while at night, the demand can fall by up to 50\% of the daily peak power. Assuming a large wind power penetration, high winds during the night can lead to excess electricity production which cannot be absorbed by the grid. In this case, ``green'' power which cannot be stored would be irrevocably lost (for storing electricity see Section~\ref{sec:storage}). Such effects can be mitigated or even vanish by the intercontinental interconnections which the Global Grid advocates. Due to time differences, when the night -- and the electricity consumption -- falls in Europe, in America it is still noon and the power consumption is at its peak. Interconnections between the two continents can take advantage of the total available power and transfer it where it is needed. In this way, the exploitation of the RES potential can reach 100\% over the whole year.

With the assumptions documented in \ref{sec:transcosts_submarine}, we estimate the cost of a 5500~km, 3~GW submarine cable to be in the range between \EUR{0.0166} and \EUR{0.0251} per delivered kWh. For a distance equivalent to the distance between Halifax, Canada and Oporto, Portugal, the cost could fall to \EUR{0.013}/kWh. Ref.~\cite{WWSglobal_grid} calculated that the costs for RES in 2020 will start from \emph{below} \$0.04/kWh and reach a maximum of \$0.13 per delivered kWh. On the other hand, the costs of conventional (mainly fossil) generation in the US will be around \$0.08/kWh and along with the incurred social costs, will total \$0.14 per delivered kWh. If we incorporate our line cost projections into these costs (in US\$\footnote{Exchange rate 2011: 1~USD=0.7119 Euros.}: \$0.023/kWh and \$0.035/kWh), it seems that, except for the most expensive RES generators, it would be more economical for the US to import RES power from Europe than operate its own fossil-fuel power plants.

%Under these circumstances, the cable could be a competitive option for participating in the daily spot auctions, where the difference between peak and off-peak prices exceeds \EUR{30}/MWh (e.g. European Power Exchange, $\langle$\url{www.epexspot.com}$\rangle$) and can surpass US\$100/MWh (e.g. PJM Interconnection [6/6/2011], $\langle$\url{www.pjm.com}$\rangle$). Day-ahead markets, although not so volatile, exhibit a peak-off-peak difference between \EUR{20-50}/MWh (depending on the area congestion), which leaves room for a profitable operation of the cable under conditions. It should be also noted, that a global grid interconnection is expected to take mainly advantage of the excess renewable energy in one area, transmitting it to areas with higher demand. This implies that prices in the low demand area could be lower or sometimes even negative, thus having a positive effect on the profitability of the line.

\subsection{Minimizing Power Reserves}
\label{sec:ancillary_services}
Power system regulations around the world require the availability of power reserves, for frequency and voltage control, in order to balance load variations and deal with contingencies. Currently, mainly generators provide such services, usually referred to as ancillary services. After the unbundling of the electricity industry, ancillary services markets have been established in several areas, from where the regional Independent System Operators (ISO) are expected to procure the necessary power reserves (e.g.~$\langle$www.swissgrid.ch; www.ercot.com$\rangle$). At the same time, however, with the increasing penetration of RES, it is expected that the amount of necessary reserves for load following and balancing purposes could increase \cite{reserves_comp}.

A higher amount of reserves is usually necessary during the day, as a high industrial and commercial activity takes place then. During the night, the load varies to a significantly lesser degree, and the need for maintaining -- and using -- reserves is lower. The existence of interconnections around the globe would help to decrease the amount of necessary power reserves within a region. For example, a significant part of the capacity withheld as control reserves from local generation sources during the day in Europe can be provided through, e.g., the interconnections with the US. During the night, the US consumes substantially less power and needs a lower amount of reserves. Thus, available control power could be provided across the Atlantic. An additional advantage is that the power sources in Europe, having more available capacity to offer, would be dispatched in a more optimal way. This would result in a decreased total generation cost and reduced power losses, which would in effect lead to a decreased electricity price for the end consumer.

It could be perhaps argued at this point that the outage of global interconnecting lines should be considered as a severe contingency and might lead to an increase in the necessary level of reserves. Indeed, the loss of a large HVDC line could exceed the loss of a large power plant in terms of power capacity. However, such interconnections are intended for connecting large interconnected systems. Taking the European power system as an example, it consists of several control areas interconnected with each other. Each area is expected to have a sufficient amount of control reserves available so that the outage of an element (e.g., a generator, a line, etc.) would have no impact on the neighboring areas \cite{ENTSOE_OpHandbook}. Equivalently, the areas interconnected by the Global Grid interconnections, e.g., the European with the American power system, should follow a similar rule. The reserves should now be withheld by, e.g., the whole European system and distributed to the individual areas. As a result, the loss of a Global Grid interconnection should have a significantly smaller effect than the loss of a power plant in a single area. At the same time, such interconnections provide an additional source of control power. As a result, significant cost savings could emerge, as the building of additional ``peaking'' gas power plants for balancing renewable energy could be avoided. This has been investigated by the authors in \cite{aboumahboub_Global_EU_grid} who compared the necessary conventional power plants in the presence or not of interconnecting lines between regions. Their results for both the European and a potential global grid showed that through interconnections the need for dispatchable conventional power plants could be reduced by two to eight times.

\subsection{Alleviating the Storage Problem}
\label{sec:storage}
Transmission grid studies already incorporate storage options in their projections for the future (e.g.,~\cite{dena_study}). Bulk quantities of storage will be necessary for absorbing non-transmissible power and relieving congestion. Although hydrogen storage and redox-flow batteries are also considered, the technologies most likely to assume this role are pump-hydro power plants and compressed-air energy storage systems (CAES) \cite{dena_study}. However, most of the available pump-hydro locations near load centers have already been exploited, while CAES has to deal with the limited number of appropriate sites and, at least for the time being, its lower efficiency ($\sim$50-70\%) \cite{dena_study}.

The HVDC links of the Global Grid have the potential to alleviate the storage problem in future power systems by absorbing excess power (i.e., with a low price) and injecting it into regions where it is needed more.  In terms of efficiency, the losses of an Ultra High Voltage DC line (e.g., $\pm$800~kV) amount to about 3\% for every 1000 km \cite{HVDC_efficiency_siemens}. This would imply that a 6000~km HVDC line with the current technology has a better efficiency than pump-hydro or compressed-air energy storage.

It could be argued at this point that, neglecting the limited availability of suitable locations, the costs for building bulk storage systems would be significantly less than the costs for a submarine intercontinental power line. In the case of hydrogen storage or redox flow battery storage, based on the costs described in \cite[p.441-444]{dena_study}, a long HVDC line is a cost-competitive option. CAES and pump-hydro power plants cost less per installed kW. Nevertheless, for an appropriate comparison, it is necessary to consider also the following three factors. In terms of capacity, an HVDC line has the capability of supplying (or absorbing) energy continuously for 8760 hours a year, as it does not need to replenish energy offered earlier. Furthermore, it can generate profits during the whole day, and in both areas that the line connects. An additional factor that needs to be taken into account is the necessity for grid reinforcements. Investigating the integration of additional RES, in \cite{dena_study}, it was shown that even if there was storage able to absorb 100\% of the excess power, 65\% of the proposed grid reinforcements would still be necessary. If the global grid interconnections could be designed in conjunction with such reinforcements, significant cost savings could arise.

Additionally, further benefits related to storage can emerge from the Global Grid concept. In~\cite{paris_longtrans}, the authors refer to untapped hydro potential in remote places in Africa, Siberia, and Alaska. Global interconnections, as also suggested in \cite{intercont_grid_hammons}, might facilitate the exploitation of these sites. Considering the significant hydropower potential in Greenland \cite{greenland_hydro}, our analysis in Section~\ref{sec:greenland} provides some further hints on such interconnections.

\subsection{Reducing the Volatility of the Electricity Prices}
Because of the intermittent nature of RES, and particularly of wind power, higher volatility of electricity prices is expected in the future \cite{dena_study}. Examples of very high prices, or its opposite, negative prices, have already been observed in power exchanges, e.g.~\cite{nicolosi_negative_prices}. The existence of several interconnections between the grids can substantially mitigate such extreme phenomena. Interconnecting lines can supply missing power in the case of high demand, or transmit excess power to where it can be absorbed. In this way, electricity variations will be minimized and the consumer can enjoy a relatively constant and, on average, lower price.

It could be argued at this point that consumers, during the night hours, will be subjected to higher prices, as electricity will be sold to high-price areas, driving also their electricity prices up. Detailed studies based on price data from existing interconnections should be carried out in order to identify more accurately the effects of this. We speculate that the mean price will fall, as the price increase during the night should be less than the decrease they will enjoy during the day (during the day expensive peak units can be replaced by the interconnections). Nevertheless, initial analyses of the NorNed link operation, an HVDC line connecting Norway and the Netherlands since 2008,  indicated no significant change in the price levels for either country, because ``capacity considerably greater than that of NorNed may be required to achieve the desired effect [i.e., a reduction of the price volatility]'' \cite{NorNed_effectOnPrice}. Such findings emphasize the need to move to a more interconnected power grid, as envisioned in the Global Grid concept.

%\subsection{Delivering the power directly to the load centers}
%An HVDC grid has the advantage that it can be superimposed over the existing AC power grid, connecting point-to-point isolated power plants with major load centers (e.g. cities or ``megacities''). Such projects have been already implemented, mainly to connect large hydro power plants to areas with large demand. Examples are the 820 km HVDC lines between Itaipu and Sao Paulo in Brazil, the 1100 km HVDC line connecting James Bay with the load centers in Canada and the U.S or the 890~km HVDC line between Three Gorges and Changzou, in China. As a result, bulk power, can be transmitted directly where it is mostly needed with less power losses and without overloading other lines or substations.

\subsection{Enhancing Power System Security}
\label{sec:security}
The Global Grid can enhance the security of the power system in several ways.

First of all, it can assist in congestion relief, thus also eliminating the congestion costs incurred by the system. Here one should, however, be careful. Such a grid, can also induce congestion in the underlying AC grid if it is not designed correctly. Ways to avoid this follow two directions. The first is to design the HVDC grid with multiple injection points, so that the power can be distributed (or absorbed) in a much wider area and then supplied to the consumers. The other direction is to reinforce currently weak network points in the underlying AC grid in order to be able to sustain the high power flows which might occur.

The Global Grid can also have a positive effect on the robustness of the network with regard to failures of transmission components. Indeed, the Global Grid builds new ``bridges'' between the power systems. A more interconnected power system leads, in general, to increased security, as new paths are created in order to serve the energy. The probability that a loss of a line will lead to an area blackout will be decreased. On the other hand, it could be argued that new couplings between networks, even at a DC level, could allow the propagation of disturbances, eventually leading to a global blackout. Measures to prevent such phenomena, even if they would only occur rarely, is a challenge to be addressed.

HVDC interconnections can also mitigate problems pertaining to the power system's dynamic behaviour. Important advantages arise from the new HVDC technologies (Voltage Source Converters), as they allow independent control of the active and reactive power. As a result, the Global Grid injection points can be foreseen to serve as reactive power providers, assisting locally in the voltage stability of the underlying grid. Furthermore, the HVDC interconnections can assist with transient stability problems. Such problems usually occur after disturbances (e.g., a short-circuit) and perturb the balance between the mechanical (input) and electrical (output) power of the machines\footnote{Machines, or synchronous machines, are often used terms in the transient stability context, denoting the generators.}. Machines with greater mechanical power than electrical power accelerate, while in the opposite case, they decelerate. The acceleration/deceleration results in a perturbation of the machines' rotating speed and a deviation from the nominal frequency. This effect is usually referred to as a loss of synchronism. If an HVDC interconnection is located near such a machine, through its active power flow control, it can absorb electrical power if the machine accelerates, or inject additional power if the machine decelerates. In this way, it could possibly prevent the loss of synchronism.

\subsection{Additional Benefits and Challenges}

An HVDC grid has the advantage that it can be superimposed over the existing AC power grid, connecting point-to-point isolated power plants with major load centers (e.g., cities or ``megacities''). As a result, bulk power can be transmitted directly to where it is most needed, with less power losses and without overloading other lines or substations.

Countries with increasing energy demand and a high carbon footprint, but with lower RES potential, could benefit from the global interconnections by importing green power. In the long run, economic benefits can emerge, as conventional fuel costs, with the addition of carbon taxes, will increase compared with the lower operating costs of the RES plants.

The Global Grid can also have an effect on increasing global cooperation, with several positive effects on the political and commercial dimensions. A significant RES potential exists in countries with developing economies. The concept proposed here can stimulate investments in these regions, and considerably assist their local economies. For example, due to the abundance of solar power in the African deserts, a significant part of this electricity can cover all the energy needs of the local population or be employed for the desalination of salt water and the mitigation of water shortage problems.

Nevertheless, the emergence of the Global Grid also creates challenges which need to be addressed. Ref.~\cite{Elliott2012} identifies issues with which a European supergrid would be confronted. Several of these also apply in the case of a global electricity network. For instance, the author refers to the fear that certain regions ``will become once again reliant on an overseas source of energy -- as with oil from the Middle East or Russian gas'', or that the HVDC links might become a target for terrorist attack. He further notes the concerns about possible political unrest in regions with abundant renewable potential. Still, the author emphasizes that electricity cannot be stored for long periods -- as with oil -- and, thus, he expects that there should be less room for conflict. In addition, he argues that the European supergrid ``would only carry a proportion of the power demand, and that, like the internet, it could have multiple network pathways'' \cite{Elliott2012}. A similar concern about the loss of a Global Grid interconnection has been addressed in Section~\ref{sec:ancillary_services}.

The multiple network pathways present an additional opportunity of the Global Grid, as they would allow for the diversification of the energy supply. Each region will be able to absorb energy from many different sources. As a result, not only the security of supply should be enhanced, but also competition between different suppliers can be established. Competition and the abundance of RES in many different world regions should diminish the strong dependence on certain countries for primary energy sources.

\section{Investments}
\label{sec:investments}
Probably one of the concerns when one envisions a global grid is its cost. The necessary infrastructure for the realization of the Global Grid involves investments in the range of billions of dollars for each interconnection. This is, however, comparable to current investments in the energy sector. Projections estimate the creation of a European offshore grid, connecting a large number of wind farms in the North Sea, at about \EUR{70}-{90} billion (\EUR{1}~billion=$10^{9}$~\EUR) \cite{Offshore_report}. The fourth generation Olkiluoto nuclear power plant in Finland has an estimated cost of about US\$4.2 billion \cite{wnn_olkiluoto}, while assumptions place the cost of the deepest offshore oil platform in the world around US\$6.7 billion \cite{forbes_perdido}. At the same time, the European Union (EU) estimates that \emph{new} electricity infrastructure will require investment costs of \EUR{140} billion until 2020 \cite{EC_Invest_Needs}. In total, the EU projects investments in the range of \EUR{1} \emph{trillion} for the European energy sector in order to ``meet expected energy demand and replace ageing infrastructure'' (\EUR{1}~trillion=$10^{12}$~\EUR)~\cite{ECgreenpaper}. To increase the investments in the transmission infrastructure is a critical priority also for the US \cite{pikeresearch_investments}. %In any case, detailed feasibility studies need to be carried out and the cost and benefits to the consumers need to be assessed. %As in the case of telecommunications, it is anticipated that big consortia from several countries, will fund the global grid projects and claim a share of the profits.

Considering the case of NorNed, in the first two months of operation NorNed generated revenues of \EUR{50}~million\footnote{NorNed is the DC link connecting Norway with the Netherlands. Data about its subsequent operation were not available. In the business plan drawn for the project, the expected revenues were initially estimated at about \EUR{64}~million per year.}, about 12\% of the invested capital \cite{norned_press}. The revenues can be translated to about \EUR{0.0556} per delivered kWh (with the same assumptions as in \ref{sec:transcosts_submarine}). Considering our calculated costs for submarine cables, and revenues similar to NorNed, the income for each delivered kWh along a 5500~km line would exceed two to four times its cost.

In the following paragraphs we will briefly describe the possible investment schemes with regard to the electricity transmission network, and subsequently, discuss the application of the investment schemes within the Global Grid environment. For a short survey on transmission investment mechanisms, the reader may refer to \cite{SCH_techreport}, and for more detailed information to e.g.~\cite{Joskow_incent_long, hogan, Joskow_Tirole_MTI}. We focus on two main investment options: regulated investment, and merchant transmission investment. We further mention additional investment alternatives, already provided for in electricity network regulations, and which could also be applicable to this concept.

\subsection{Investment Mechanisms}
Two main investment options exist with respect to transmission infrastructure: the regulated investment, which is the most common kind of transmission investment, and the merchant transmission investment, which emerged after the restructuring of the electric power industry. In Europe, DC interconnections such as NorNed and BritNed (which connects the UK with the Netherlands) seem to follow the merchant transmission investment model. A mix of these two investment schemes, with part of the capacity subject to regulation and the rest available for generating profits from trade, can also be anticipated. Such an option is already provided for in the EC regulations (see \cite{EC_2003}, Art.~7.(4b)).

\subsection{Investments in the Global Grid}
We anticipate that most long intercontinental cables will be implemented under a regulated investment regime. Based on the analysis in Section \ref{sec:smoothing}, global interconnections for transmitting RES power seem a cost-competitive option. Therefore, and in conjunction with the increased amount of renewables they facilitate, we expect that they should increase social welfare and be eligible for a regulated investment. Furthermore, regulated investments in long cables may be more favorable than merchant investments on two accounts. First, the capital intensive nature of these interconnections induces a significant amount of risk that a private consortium might not undertake. At the same time, the expected revenues from electricity trade in comparison to the costs per delivered MWh of a long submarine cable (e.g. 5500~km), although potentially higher, may not provide for an attractive private investment, in a first phase.

For instance, based on the cost projections of \ref{sec:transcosts_submarine}, the cable could be a competitive option for participating in the daily spot auctions, where the difference between peak and off-peak prices exceeds \EUR{30}/MWh (e.g., European Power Exchange, $\langle$\url{www.epexspot.com}$\rangle$) and can surpass US\$100/MWh (e.g., PJM Interconnection [6/6/2011], $\langle$\url{www.pjm.com}$\rangle$). Day-ahead markets, although not so volatile, exhibit a peak--off-peak difference between \EUR{20-50}/MWh (depending on the area congestion), which leaves room for a profitable operation of the cable under conditions. A long-term contract though, for such a long line, would probably not generate any profits under current circumstances. It could therefore be expected that a private initiative may require higher profit margins, or less risk, in order to undertake such an investment.

With the increasing integration of renewable energy sources, which essentially have a near-zero marginal cost, it can be argued that the opportunities for price arbitrage could disappear. Indeed, this could be true for long-term contracts and day-ahead markets. However, as uncertainty for the delivered energy increases due to the intermittency of RES, larger energy volumes will be traded in the highly volatile intra-day spot markets. Therefore, it is expected that a significant amount of the line's profit will result from arbitrage in intra-day spot markets.

Merchant investors could successfully launch smaller-scale projects, which would not only complement but also facilitate the Global Grid (e.g., inter-regional HVDC interconnections up to 2000~km). For example, assuming that state capital will fund an interconnection between Europe and North America, independent initiatives could invest in an HVDC line between the UK and France, or between France and Italy. These smaller interconnections, on the one hand, could benefit from the existence of the intercontinental line by trading power from North America to Central Europe and the reverse. On the other hand, the same interconnections would further allow the Global Grid to expand, since interconnections from Italy with Africa and the Middle East would be facilitated. As power flows may have more than one alternative route to follow, we expect that competition between the lines will exist.

Here it should be noted that an increasing number of interconnections could reduce the price volatility and, thus, discourage additional merchant investors. On this account, according to \cite{eon_uk}, regulated investments are expected to also be necessary for the European supergrid, mentioning that ``the incentive for a merchant investment will decrease not allowing the full benefits from an integrated European grid to be realized.'' On the other hand, studies have shown that the NorNed interconnector had no significant effect on prices during its first two years of operation, concluding that a substantially larger capacity might be necessary to achieve reduction in price volatility \cite{NorNed_effectOnPrice}. Concerning the long submarine interconnections of the Global Grid, we find the argument about the reduction in price volatility reasonable, although this effect would possibly be more difficult to occur due to the magnitude of the networks.

An option for the Global Grid would also be a mix of regulated with merchant investment, or alternatively, a subsidy scheme. This is the case for the NordBalt interconnection, between Sweden and Lithuania. The European Union supports the project, with \EUR{131}~million, which compares to its total budget of \EUR{550}~million \cite{NordBalt_totalcosts}. Similar subsidies, scaled up with respect to the budget, can be envisioned for the Global Grid interconnections.

Literature concludes that developing good regulatory mechanisms which will also provide opportunities for merchant investors to develop projects seems to be a good solution, but at the same time a significant research challenge \cite{Joskow_Tirole_MTI, Brunekreeft_Cambridge}. The regulation -- or non-regulation -- of the Global Grid investments remains equally a challenge.

\section{Operation}
\label{sec:operation}

Besides competition, both within the region and between the lines, interconnectors could facilitate the establishment of market couplings, eventually leading to a common global market environment. For instance, NorNed and BritNed in the European region, along with the existing interconnections, led to the coupling of the Nordic and the UK electricity market to Central Europe, forming and expanding the Central West European electricity market (CWE, see $\langle$www.casc.eu$\rangle$). The NordBalt interconnection aims at connecting also the Baltic Market with the CWE \cite{NordBalt_cable}.

In the following paragraphs we will outline possible operation schemes of the Global Grid.

\subsection{Organization of a Global Market Environment}
Within the Global Grid context, we anticipate the need for having the activities between the regional regulators coordinated by a common coordinating body. We envision the establishment of a new body, which we will call the ``Global Regulator'', that will assume a supervisory role. Its main responsibilities would be to provide a forum for communication among the interested parties, coordinate investments, and ensure a competitive market environment.

During the construction of the first interconnections, the Global Grid could operate in a more decentralized way, either with explicit auctions, or coupling two markets through an implicit auctioning system\footnote{For more information about capacity allocation mechanisms, refer to \cite{krause_phd, kurzidem_phd}}. However, as soon as the Global Grid acquires a more meshed form, we anticipate that conditions would be more favorable for centralizing the coordination of electricity trade. We identify two possible options for the organization of a Global Electricity Market. In both cases, we envision the establishment of a new regulating body, the ``Global System Operator''. The Global System Operator could have similar responsibilities to the currently existing Independent System Operators.

The first option we could envision is a hierarchical market model, where we distinguish the DC grid from the underlying regional AC networks. A Global Market can be formed in which each regional market participates as an individual player. The Global System Operator, in charge of the global market clearing, will clear the market and assign the corresponding amounts of power to each player. Here, considerations should be made on a fair and transparent way that the HVDC flows can be calculated. Then each regional market, assuming the injection/absorption of the HVDC links as given, will receive the bids of the local generators and consumers and clear its own market under the local constraints. The advantage of such a scheme is that the structure of each regional market does not need to be significantly modified. An alternative is a more ``horizontal'' operation of the global market. A Global Power Exchange could be established to facilitate power trade. Entities around the globe will be able to participate in this global market and offer bids for power supply or demand. In this case, the induced power flows will not be taken into account during the market clearing, thus keeping the complexity of such a scheme low.

A feature of the Global Grid could also be the participation in ancillary services markets, allowing its users to place bids for offering active or reactive power\footnote{HVDC lines equipped with Voltage Source Converters are able to control the active power independently from the reactive power. Reactive power is necessary in order to deal with voltage instability, which is usually a local phenomenon requiring local actions. Therefore, the provision of reactive power will most probably be limited to the market area to which the HVDC injection point belongs; by comparison, the active control power could, under certain assumptions, be offered to regions different from the injection point of the HVDC line.} in the regional ancillary services markets. In this context, we also envision the establishment of a Global Ancillary Services Market. Such functions can be facilitated if the HVDC line buys control power from another control market, in the sense of arbitrage.

%Studies need to be carried out about the effects of an ineffective regulation practices on the global market operation.

\subsection{Challenges in Operation}
Considering the operation of a globally interconnected HVDC grid, with several actors and owners, in this paragraph we identify issues that would need to be addressed if we decide to move towards a Global Grid. For example, a market model needs to be (re)designed for the coupling of the Global Grid with the underlying AC grid. How would the pricing in the multi-terminal nodes take place and how would the capacity of multiple parallel HVDC links be allocated? A harmonization of the regional markets or, alternatively, the coupling of the different market structures should also be investigated. Furthermore, the oligopolistic behaviour of the power producers in a global market environment should be studied: will they retain market power or is the Global Grid leading closer to perfect competition? In a similar sense, are the HVDC link owners going to attain market power? Different regulations might need to be developed and tested in order to identify one which will not permit monopolistic/oligopolistic behaviour on the part of merchant investors.

Concerning the market operation with increased RES integration, a more detailed study about price fluctuations needs to be carried out. Current market designs exhibit inefficiencies as more renewable production with near-zero marginal cost participates in the market operations. Peak and off-peak prices may not be driven by the demand fluctuation in the future, but rather by the availability of renewable generation. Different price patterns may emerge, which will probably be dependent on the residual demand\footnote{The residual demand is defined as the actual demand minus the renewable energy supply.}. Still, global interconnections can increase the reliability of the energy supply, and benefit significantly from fluctuations in the residual demand between the regions they connect.

Studying the effect of the Global Grid on security is an equally important aspect. Methods that will adapt the current control and security mechanisms in order to take advantage of the overlying HVDC grid should be considered. For example, it needs to be investigated whether the provision of control power from the HVDC line will be necessary for the operation of the Global Grid. At the same time, the possibility of a global blackout, its effects and methods to prevent it, should also be studied.

\section{Discussion}
\label{sec:discussion}
Alternatives of the Global Grid for the global energy supply of the future can also be envisioned.

A reasonable assumption would be to continue with the ``business as usual'', where fossil fuels and nuclear energy will account for the majority of the energy supply. Nevertheless, increasing environmental awareness, the approaching decline of world oil production (peak oil), increasing prices, and the concerns for oil and gas security of supply are changing the current paradigm and will probably lead to increased penetration of renewable energy sources.

A different alternative would be to focus solely on dispersed generation power plants, following the current trend. This alternative, among several other advantages, alleviates the need for investments in transmission lines, allowing the energy to be locally produced and consumed. Although it is widely assumed that due to the economies of scale, the cost of investing in such renewable energy technologies will continuously fall, in \cite{Dinica} it is suggested that in the long run a flattening out or a U-shaped curve for production costs would be more likely.  According to the author, ``[in Spain] the increases in technological performances and installed capacity per turbine did not compensate for the decrease in resource quality and availability'', and this resulted in an increase of the windpower production costs from \mbox{5.1-6.5}~\euro c/kWh in 2005 to an estimated \mbox{5.6-7.0}~\euro c/kWh in 2010. A reasonable assumption, following this argument, would be that a break-even point will exist in the future, where tapping energy from remote locations with abundant resources, would be competitive with new RES installations near the load centers. Therefore, we would expect that harvesting these remote resources would complement and reinforce the efforts currently being taken towards a more ``green'' energy future.

Undoubtedly, a global electricity grid is not the sole option for covering the global energy demand in the future. Several new gas pipelines have recently become operational or are in planning phase in Europe, Asia and elsewhere (e.g., NordStream $\langle$\url{www.nordstream.com}$\rangle$, SouthStream $\langle$\url{www.south-stream.info}$\rangle$, Central Asia-China \cite{centralasia_china}), as natural gas is becoming increasingly important in the global energy mix. In a high-RES energy future, mixing hydrogen with natural gas could be envisioned as alternative for such a network. The main advantage of this option is that gas can be stored for longer periods than electricity, and, thus, provide a non-intermittent source of energy. Although at greater sea depths it is questionable if the construction of gas pipelines is more favorable than submarine cables, still, the option of a global gas pipeline network which could primarily facilitate the transport of ``green'' energy could also be investigated.

Assuming the need to harvest significant amounts of energy from remote locations will exist, an alternative energy carrier, as already mentioned, could be hydrogen. For example, ``green'' electricity could electrolyze water and produce hydrogen. Hydrogen could then be transported by conventional means and transformed back to electricity by fuel cells near the load centers. However, at the moment, the round-trip efficiency of such a scheme is relatively low (about 32\% according to \cite{dena_study}).

Thus, we expect that the Global Grid will be one of the most probable alternatives, as we move towards an ever increasing share of RES in the electricity production.

%The size of the necessary investments is probably the biggest concern at the moment. We claim, though, that the vision of the Global Grid today can be the reality of the future. Larger interconnections are already planned and built \cite[e.g. European SuperGrid,][]{eon_uk}, while tapping renewable energy from remote sources is already in the planning phase \cite[e.g.][]{Desertec}. The Global Grid is simply the next logical step.

The first interconnections have already started to be constructed independently (e.g., NorNed, BritNed, etc.), focusing on the benefits and impacts at a regional level. Several line reinforcements will be necessary in any case, for the integration of new RES plants in every region. We argue that it will be beneficial to consider synergies with a possible global grid even in the planning stages. As the network of HVDC lines expands and connects larger regions, regional power systems must start to explore in more detail the opportunities a meshed global HVDC grid can offer.

\section{Conclusions}
\label{sec:conclusions}
The Global Grid advocates the connection of all regional power systems into one electricity transmission system spanning the whole globe. Power systems currently are forming larger and larger interconnections, while ongoing projects plan to supply, e.g., Europe with ``green power'' from the North Sea or the African desert. Environmental awareness and increased electricity consumption will lead more investments towards renewable energy sources, abundant in remote locations (off-shore or in deserts). The Global Grid will facilitate the transmission of this ``green'' electricity to the load centers, serving as a backbone. This paper has focused on the introduction of the concept from a technological point of view and has highlighted the multiple opportunities that emerge. Studying the possibility of a submarine cable connecting Europe with North America, we showed that, except for the most expensive RES generators, it would be more economical for the US to import RES power from Europe than to operate its own fossil-fuel power plants. This paper has also briefly discussed  investments and market operation in a Global Grid environment.

Concluding the paper, we argue that the first working groups to examine in more detail several different aspects of the Global Grid should start to be formed. Studies could be carried out on a technical, economical, and societal level in order to substantiate the benefits and the challenges that need to be addressed. More detailed feasibility studies, for different world regions, will need to be carried out. The risks that such projects would entail need also to be quantified and incorporated in the studies. The research community could also actively participate in order to tackle the emerging problems and develop the necessary methods and tools that could facilitate the Global Grid.

\section{Acknowledgements}
This paper was not funded by any interest group, company, or government agency. The authors would like to thank Prof. Louis Wehenkel, Dr. Thilo Krause, and the reviewers for their helpful comments.

\appendix
\section{Cost of a Long-Distance Submarine Cable}
\label{sec:transcosts_submarine}
In this section we attempt a rough estimation of the costs for a submarine HVDC cable. Several studies have already tried to identify such costs. In \cite[p.~133]{Hauth_HVDCcosts} the cost of an HVDC 1000~MW, $\pm$400~kV, 100~km bipolar cable were estimated at around US\$126.7~million, which is about \EUR{1.36}~million/km in 2007 prices\footnote{The year 2007 was selected for consistent comparison with the rest of the studies and the implemented projects (e.g. NorNed, SAPEI and BritNed were launched or commissioned between 2006-2009). Exchange rate: 1~USD=0.8587~EUR in 4/1997 (\url{www.x-rates.com}) and average Euro inflation rate of 2.24\% from 1991 to 2010 (\url{http://www.tradingeconomics.com/euro-area/inflation-cpi})}. Regarding the converters, the authors (p.~66) suggest that the costs of a $\pm$350~kV, 1000~MW terminal should be equal to \EUR{115}~million and of a $\pm$500~kV, 3000~MW terminal to \EUR{242}~million~(2007 prices). Consistent with that, in \cite{Weigt_eurosupergrid} terminal costs in the range between \EUR{150} to \EUR{350}~million are assumed (onshore or offshore terminal), but the authors mention that submarine cables can cost ``up to'' \EUR{2500}~million per 1000 km. Their assumptions are based on \cite{DLR_HVDCcosts}. More specifically, in \cite{DLR_HVDCcosts} (p.~24) the costs of a 5000~MW HVDC sea cable are estimated to be about \EUR{2500}~million/1000~km when operating at $\pm$600~kV and \EUR{1800}~million/1000~km when operating at $\pm$800~kV. The costs of the respective terminal converters are estimated to be about \EUR{250-350}~million. In \cite{WWSglobal_grid}, the authors made a detailed analysis about the HVDC overhead line costs. Based on this analysis, they subsequently outlined low-, mid- and high-cost cases for HVDC cables, with costs ranging from \EUR{0.81}~million/km to \EUR{4.17}~million/km for a 3000~MW HVDC submarine cable\footnote{Exchange rate (2007): 1 USD=0.6808 EUR.}. The costs for the converters were in the range of \EUR{200-310}~million. In \cite{Iceland_UK}, costs in the range of \EUR{419-456}~million/1000~km are mentioned\footnote{2007 prices. 1 GBP=1.40219 EUR in 1992. Average Euro inflation rate of 2.24\%. } including transportation and laying for a $\pm$450~kV, 600~MW cable, while the costs for a $\pm$450~kV, 1200~GW terminal converter are assumed equal to \EUR{301}~million.

\begin{table*}[!htb]
\centering
\begin{threeparttable}[htb]%\vspace{-0.3cm}
\caption{Costs of HVDC submarine projects}
\label{tab:HVDCprojects_cost}
\begin{tabular}{rrrrrp{4.6cm}}
%\begin{tabular}{rrrrrl}
  \hline
  % after \\: \hline or \cline{col1-col2} \cline{col3-col4} ...
  NorNed & SAPEI & BritNed & NorGer & NordBalt & \\
  \hline
  $\pm$450 & $\pm$500 & $\pm$450 & 450-500 & $\pm$300 &Voltage (in kV, DC side)\\
  700 & 1000 & 1000 & 1400 & 700 & Capacity (MW)\\
  580 & 435 & 260 & 570 & 450 & Length (km)\\
  410 & 1600 & n/a & 410 & n/a & Maximum Sea Depth (m) \\
  \EUR{600}\tnote{a} & \EUR{750}\tnote{b} & \EUR{600}\tnote{c} & \EUR{1400} ($\pm$30\%)\tnote{d} & \EUR{550}\tnote{e} & Cost (in millions)\\
  \EUR{300}\tnote{g} & \EUR{450}\tnote{g} & \EUR{300}\tnote{g} & \EUR{680-1520}\tnote{g} & \EUR{270}\tnote{f} & Cost without converters\tnote{h} (in millions)\\
  \EUR{0.52} & \EUR{1.03} & \EUR{1.15} & \EUR{1.19-2.67} & \EUR{0.60} & Cost/km (in millions) \\
  \hline
\end{tabular}
\begin{tablenotes}[para]
\item[a]{\cite{NorNed_costs}}
\item[b]{\url{www.sapei.it}}
\item[c]{\cite{BritNed_costs}}
\item[d]{planning phase; \cite{NorGer_wiki}}
\item[e]{planning/construction phase; \cite{NordBalt_totalcosts}}
\item[f]{both for cable supply and installation; \cite{NordBalt_cable}}
\item[g]{The cost of each terminal converter was \emph{assumed} equal to \EUR{150}~million, based on the lower limit of the mentioned studies, although one would expect that the cost of the NorGer terminal stations could be somewhat higher.}
\item[h]{We assume that these costs correspond to the manufacturing and \emph{installation} of the cable.}
\end{tablenotes}
\end{threeparttable}%\vspace{-0.3cm}
\end{table*}
%possibilities for \begin{tablenotes}[para], or [normal] or [flushleft] or [online]

Table \ref{tab:HVDCprojects_cost} presents the costs of five existing (and planned) submarine HVDC links. NorNed is the longest submarine cable to date, connecting Norway with the Netherlands. SAPEI is currently the deepest submarine cable in the world, connecting Sardinia with mainland Italy. BritNed connects the UK with the Netherlands and has been in operation since April 2011. NorGer, connecting Norway with Germany, is still in the planning phase, with estimated project completion in 2015. Also planned is the NordBalt cable, between Sweden with Lithuania, connecting the Baltic electricity market with the Nordic and the European market. Completion of the NordBalt cable is projected for 2015/2016.

We observe that the projections in \cite{Hauth_HVDCcosts}, although slightly overestimated, are in line with the calculated costs of existing projects. Consistent with the costs of NorNed are also the assumptions in \cite{Iceland_UK}. For higher capacities, the authors assume parallel lines of 600 MW, which brings them in line with SAPEI and BritNed, although with slightly underestimated costs. It should be, nevertheless, noted, that they have assumed higher costs for the converters than we did in Table~\ref{tab:HVDCprojects_cost}. Comparing the $\pm$600~kV option of \cite{DLR_HVDCcosts} with Table~\ref{tab:HVDCprojects_cost}, which have a similar voltage level, we note that the costs are overestimated. But it should be taken into account that in \cite{DLR_HVDCcosts} the authors refer to costs for a 5000~MW line. It is also worth mentioning that in the NordBalt link, where the cable cost was known, the cost of the \emph{two} converters could be estimated at around \EUR{280} million, similar to our assumptions\footnote{In \cite{NordBalt_converters}, nevertheless, it is mentioned that the total cost of the two converters was \EUR{147} million, with the remaining \EUR{130} million being -- obviously -- allocated for other costs.}.

With these considerations in mind, we distinguish two cases for our calculations. In both cases we assume a 3000~MW, $\pm$800~kV submarine cable with a length of 5500~km. Note that the distance from Halifax, Canada to Oporto, Portugal is 4338~km, while the distance from New York City to Oporto is 5334~km. We selected the $\pm$800~kV option, as we believe that for long-distance transmission, higher voltage levels will be adopted. As a high-cost case, we assume a cost of \EUR{1.8}~million/km for our 3000~MW line, the same as what was suggested in \cite{DLR_HVDCcosts} for a 5000~MW sea cable. As a low-cost case, we assume the maximum cost of the completed HVDC projects in Table~\ref{tab:HVDCprojects_cost}. This is \EUR{1.15}~million/km. The rest of the cost assumptions will be the same for both cases.

Due to the higher voltage and the large capacity of the line, we assume the cost of each terminal converter to be \EUR{300}~million. Regarding the line thermal losses, in \cite{DLR_HVDCcosts} 2.5\% per 1000~km is suggested, while in \cite{HVDC_efficiency_siemens}, it is mentioned that it is less than 3\% for a similar line. We select a value closer to the upper limit, i.e., 3\% per 1000~km, while each of the terminals has additional losses of 0.6\% \cite{Iceland_UK, Negra_HVDCconverter_efficiency}. In our calculations we also incorporate the availability of the line due to unscheduled outages, equal to 99\%, as suggested in \cite{ABB_HVDCavailability}\footnote{The authors in \cite{Iceland_UK} assume an availability of 92.2\% to 96.4\%; however, significant advances in cable technology have taken place during the last two decades.}. In a memo during the design of the NorNed cable, it was concluded that the ``technical lifetime expectancy [is] \emph{in excess} of 40 years for an HVDC system with submarine cables'' \cite{HVDC_lifetime}. Consistent with that, NorGer \cite{NorGer_wiki}, as well as Ref.~\cite{Iceland_UK}, assume a similar lifetime for their project. Our assumptions follow the same line, adopting a life expectancy of 40 years. Additionally, a 3\% discount rate was assumed, as recommended by \cite[p.~9]{NREL_discountrate}. %\footnote{\cite{OMB_discountrate} suggest to carry out the analyses with both 3\% and 7\% discount rates. The selected discount rate is probably more appropriate for regulated investments, funded by state capital. For merchant investments, a higher discount rate would be more suitable.}
%\footnote{The selected discount rate lies within the range suggested by \cite{OMB_discountrate}, i.e.~\mbox{3-7\%}, and is probably more appropriate for regulated investments, funded by state capital. For merchant investments, a higher discount rate would be more suitable.}.
The last parameter we take into account is the number of hours per day that the line can operate at full capacity. In the auction rules of NorNed \cite{Norned_auctionrules}, the ``Ramping Constraint Interval'' is defined. According to this, the available transfer capacity is decreased to 300~MW (i.e., about half of NorNed's nominal capacity) one hour before and one hour after each change of the power flow direction. In our calculations, we assumed that every day two changes in the power flow direction take place and, as a result, for four hours per day the line will operate at 50\% of its rated capacity. It should be noted, that from a technical point of view, the ``Ramping Constraint Interval'' is necessary only for the Current Source Converter technology (CSC-HVDC), as with NorNed. In order to change the power flow direction in the CSC-HVDC, the voltage polarity must be inverted. This limitation does not exist in the Voltage Source Converter HVDC technology (VSC-HVDC), and is one of the reasons why VSC-HVDC lines are considered more suitable for an HVDC grid. Although the Global Grid interconnections will probably be based on the VSC-HVDC technology, we include this constraint in our calculations, thus assuming a utilization factor of the line of about 83\%.

According to our calculations, in the low-cost scenario, which again is in the upper cost limit of the already implemented projects, each delivered MWh has a cost of \EUR{0.0166}/kWh. In the high-cost scenario, this cost rises to \EUR{0.0251}/kWh.
%Taking into account that the difference between peak and off-peak spot prices exceeds \EUR{30} each day (e.g European Power Exchange $\langle$\url{www.epexspot.com}$\rangle$; PJM Interconnection $\langle$\url{www.pjm.com}$\rangle$), the line seems to be a competitive option in the current market environment\footnote{PJM real-time peak and off-peak price difference can exceed US\$100/MWh (e.g 6/6/2011). Day-ahead markets are not so volatile with peak--off-peak difference lying between \EUR{20}-\EUR{50}/MWh, depending on the area congestion. Long-term contracts exhibit a price difference of \EUR{10}/MWh, not allowing for profit under the current cost assumptions.}.
For a length of 4400~km, as would be the distance between Halifax and Oporto, the cost falls down to \EUR{0.013}/kWh. %Here, two more remarks should be made. First, a Global Grid is expected to take mainly advantage of the excess renewable energy in one area, transmitting it to areas with higher demand. This implies that prices in the low demand area could be lower or sometimes even negative, thus increasing the profitability of the line. Furthermore, \cite{WWSglobal_grid} calculated that the costs for RES in 2020 will start from below \$0.04/kWh and reach a maximum of \$0.13 per delivered kWh. On the other hand, the costs of conventional (mainly fossil) generation in the US, will be around \$0.08/kWh and along with the incurred social costs, will total \$0.22 per delivered kWh. If we incorporate our HVDC cable cost projections in these costs (in US\$\footnote{Exchange rate 2011: 1~USD=0.7119 Euros.}: \$0.023/kWh and \$0.035/kWh), it seems that it would be more economical for the US to import RES power from Europe than to operate its own fossil-fueled power plants.

%Our findings are in line with similar studies in the literature. \cite{paris_grandinga} found that it is economically competitive to build long transmission lines in order to transfer hydropower electricity from Central Africa to Italy than to operate Italian fossil fuel power plants. \cite{Iceland_UK} carried out a detailed feasibility study and concluded that imported RES power from Iceland to the UK is competitive with conventional fossil plants in England and Wales ``in all but the very worst scenarios''. \cite{Czisch_eurosupergrid} draws similar conclusions about the European region, concluding that interconnected dispersed RES generation in Europe, North Africa and West Asia could supply power in competitive prices with the European fossil-fuel power plants.

\section{Transmission Costs for a Wind Farm in Greenland}
\label{sec:transcosts_greenland}
Continuing with our analysis from \ref{sec:transcosts_submarine}, we intend to roughly estimate the transmission cost for a wind farm in Greenland. To this end, we assume that a 3~GW wind farm on the eastern shores of Greenland is feasible, and, after a feasibility analysis, the investors have decided to connect the wind farm through Iceland and the Faroe islands to North UK. Here, we extend this analysis, in order to investigate whether it would be profitable for the investors to construct at the same time an additional transmission line to North America.

This project, except for cables, will also involve HVDC overhead lines. The authors in \cite{WWSglobal_grid} made a detailed study and concluded that the costs for a $\pm$450~kV, 3000~MW HVDC overhead line lie in the range of about \$0.3~million/km to \$2.0~million/km (or \EUR{0.21}--\EUR{1.42}~million/km). In \cite{Weigt_eurosupergrid}, the cost of overhead lines is estimated at \EUR{250-450}~million/1000~km based on studies of existing projects (e.g., Three Gorges in China), but due to NIMBY\footnote{``Not In My Back Yard''} problems, the authors assume a cost of \EUR{600}~million/1000~km. We will assume the same cost in our study, which also corresponds to the mid-cost estimates of \cite{WWSglobal_grid} [i.e., \$842~million/1000~km]. The rest of the line parameters are assumed equal to those of the cable alternative, as described in \ref{sec:transcosts_submarine}, while for all the cables and overhead lines we assume a capacity of 3000~MW. The capacity factor of the off-shore wind park is assumed to be 40\%, as in \cite{WWSglobal_grid}.

\begin{table*}[!hptb]
\centering
\begin{threeparttable}[htb]%\vspace{-0.3cm}
\caption{Length of transmission line segments connecting the wind farm in Tasiilaq, Greenland with North UK and Quebec City in North America\tnote{b}.}
\label{tab:Greenland_translength}
%\begin{tabular}{lrl|lrl}
\begin{tabular}{p{2.7cm}rl|p{2.7cm}rl}
  \hline
  \multicolumn{3}{c|}{\emph{Greenland to UK}} & \multicolumn{3}{|c}{\emph{Greenland to Quebec City}}\\
   From--To & Length (km) & Type & From--To & Length (km) & Type\\
  \hline
  Tasiilaq--Reykjavik & 770 & Sea Cable & Tasiilaq--Nuuk & 667 & OHL\tnote{a} \\
  Reykjavik--Reydarfjördur & 387 & OHL\tnote{a} & Nuuk--Auyuittuq Nat.~Park & 550 & Sea Cable \\
  Reydarfjördur--Faroe Islands & 452 & Sea Cable & Auyuittuq Nat.~Park--North Quebec & 510 & Sea Cable \\
  Faroe Islands--North UK & 457 & Sea Cable & North Quebec--Quebec City & 1542 & OHL\tnote{a} \\
  \hline
\end{tabular}
\begin{tablenotes}[para]
\item[a]{Overhead Line.}
\item[b]{Quebec City is probably supplied with ``cheap'' hydropower from James Bay. The distances to load centers such as New York City or London were similar (i.e. 714~km Quebec City--New York City and 844~km North UK--London) and respective overhead lines were not considered in the study. The interconnecting lines and cables are assumed straight lines. Distances were measured by $\langle$www.freemaptools.com$\rangle$.}
\end{tablenotes}
\end{threeparttable}%\vspace{-0.3cm}
\end{table*}
%possibilities for \begin{tablenotes}[para], or [normal] or [flushleft] or [online]

Table \ref{tab:Greenland_translength} presents the calculated distances for the two transmission paths to North UK in Europe and Quebec City in North America. Observe that the length of the submarine cables is not substantially longer than currently existing projects, while the sea depth on no occasion exceeds 1500~m (see also Fig.~\ref{fig:map_elevation}). In total, we assumed four terminal stations, two of which are located in Tasiilaq, Greenland.

The cost for transmitting power from Greenland to the UK is estimated to be around \EUR{0.014}/kWh in the low-cost scenario and \EUR{0.019}/kWh in the high-cost scenario. In case the wind farm owner now decides to additionally construct a transmission line to North America, the cost for delivering the produced power rises to \EUR{0.029}/kWh in the low-cost scenario and \EUR{0.038}/kWh in the high-cost scenario. Considering the lower limit of cost projections for producing electricity from off-shore wind in \cite{WWSglobal_grid} [i.e., \EUR{0.06}/kWh], we find that the cost of a delivered kWh increases between 21\% and 25\% if both transmission paths are constructed. However, when the wind farm is connected to both continents, it can take advantage of the time zone diversity and always sell at peak price. Assuming that the wind probability distribution has no correlation with the time of the day, we expect that 50\% of the produced energy will be directed to North America and the other 50\% to the UK and Europe. If off-peak prices are half of the peak prices, the advantage of the wind park translates to 33\% increased revenues\footnote{In our case study the increase in revenues would be 31\% due to the increased losses in the path Greenland-Quebec City (longer transmission line).}. It seems that an investment in an additional line could be profitable for the wind farm owner in such a case.

In addition, however, to the transmission of wind power, a line that connects the UK with North America allows trading electricity. We found that the wind farm delivers 4822 GWh/year to the UK and 4637 GWh/year to North America. The capacity of the line is such that it can accommodate an additional flow of 10095 GWh/year. Assuming that the transmission line allocates the unused capacity to electricity trade, it can deliver in total 19554 GWh/year. The transmission costs in this case fall between \EUR{0.014}/kWh and \EUR{0.0185}/kWh. Observe that these costs are similar to the transmission costs if the wind farm was connected only with the UK. At the same time, the revenues increase significantly, since not only is the wind energy always sold at peak price, but also additional income results from the allocation of line capacity to electricity trade.

As a conclusion for both cost assessment sections, it should be noted that our calculations serve as rough estimates. More detailed analyses should be carried out, with better estimates on costs, projections for future electricity prices and their volatility (which could possibly be higher), and incorporating parameters such as landscape and seabed terrain for the construction of the lines and the laying of the cables.

\section{Maps}
\label{sec:maps}
\begin{figure}[!hptb]
    \centering
    \includegraphics[width=1\columnwidth]{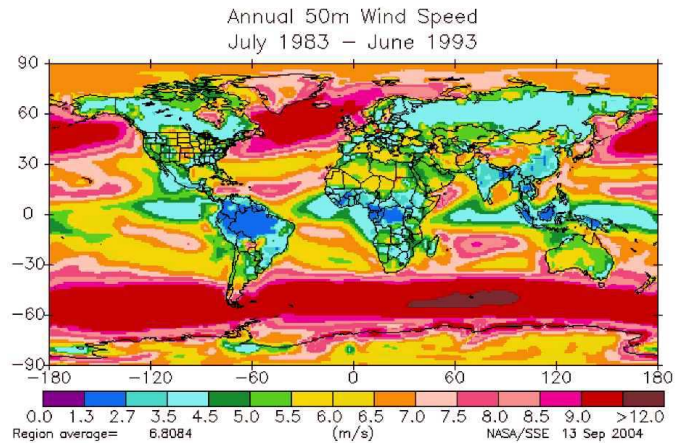}
    \caption{Wind Potential -- wind speed 50~m above earth surface. (Source: NASA/SSE) [In \cite{WWSglobal_generation}, a similar map for wind speeds 100~m above earth surface is presented]}\label{fig:map_windpotential}
%    \vspace{-0.3cm}
\end{figure}

\begin{figure}[!hptb]
    \centering
    \includegraphics[width=1\columnwidth]{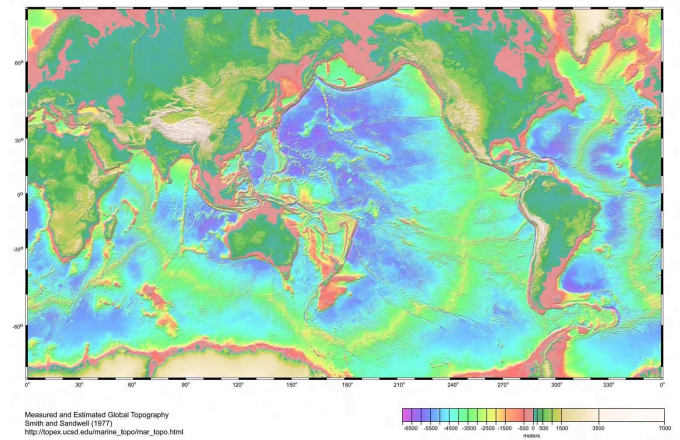}
    \caption{Earth elevation (Source: Smith and Sandwell (1977))}\label{fig:map_elevation}
\end{figure}

\begin{figure}[!hptb]
    \centering
    \includegraphics[width=1\columnwidth]{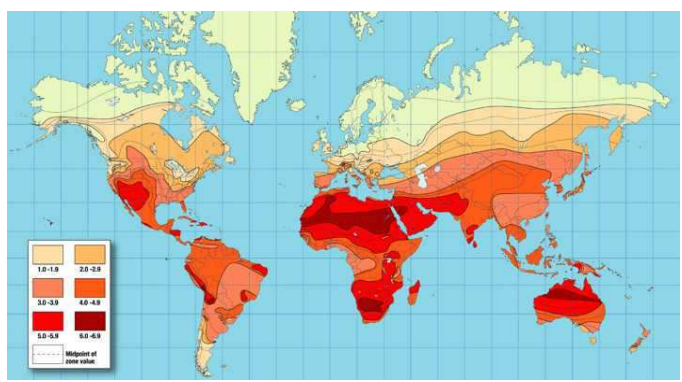}
    \caption{Solar Potential (Source: http://www.oksolar.com/) } \label{fig:map_solarpotential}
\end{figure}

The following figures present the maps, according to which the siting of the RES power plants in Fig.~\ref{fig:GlobalGrid_withExisting} has been assumed. As shown in Fig.~\ref{fig:map_windpotential}, off-shore areas experience in general higher average wind speeds than land regions. Areas in the North Atlantic and Pacific Oceans, as well as a belt off the Antarctica coast demonstrate the highest average wind speeds. Comparing Fig.~\ref{fig:map_windpotential} with Fig.~\ref{fig:map_elevation}, it is possible to identify regions where sea depths might permit the installation of off-shore wind farms. Undoubtedly, more detailed analysis is necessary in order to quantify the actual electric energy output at each site, also taking into account the wind turbine rated speed. In Fig.~\ref{fig:map_solarpotential}, the solar potential of different regions is presented. As it would be expected, areas in the tropical zone experience the highest solar irradiation. Deserts such as the Sahara in Africa, the Mojave in the USA, or the Thar Desert in India, among others, seem to demonstrate significant potential for concentrated solar thermal or large-scale solar PV applications (e.g., \cite{kurokawa_vlspv}).

\bibliographystyle{model3-num-names}
\bibliography{conceptpaper_biblio}

\begin{thebibliography}{67}
\providecommand{\natexlab}[1]{#1}
\providecommand{\url}[1]{\texttt{#1}}
\providecommand{\urlprefix}{URL }
\expandafter\ifx\csname urlstyle\endcsname\relax
  \providecommand{\doi}[1]{doi:\discretionary{}{}{}#1}\else
  \providecommand{\doi}{doi:\discretionary{}{}{}\begingroup
  \urlstyle{rm}\Url}\fi
\providecommand{\eprint}[2][]{\url{#2}}
\providecommand{\BIBand}{and}
\providecommand{\bibinfo}[2]{#2}
\ifx\xfnm\undefined \def\xfnm[#1]{\unskip,\space#1}\fi
%Type = Misc
\bibitem[{{EC}(2008)}]{EC_202020}
\bibinfo{author}{{EC}\xfnm[]}.
\newblock \bibinfo{title}{Impact assessment on the {EU}'s objectives on climate
  change and renewable energy}.
\newblock \bibinfo{howpublished}{European Commission}; \bibinfo{year}{2008}.
%Type = Misc
\bibitem[{{State of California}(2011)}]{California_energybill}
\bibinfo{author}{{State of California}\xfnm[]}.
\newblock \bibinfo{title}{Senate {B}ill {X}1-2 (sbx1 2)}.
\newblock \bibinfo{year}{2011}.
%Type = Misc
\bibitem[{{dena}(2010)}]{dena_study}
\bibinfo{author}{{dena}\xfnm[]}.
\newblock \bibinfo{title}{{d}ena {G}rid {S}tudy {II} | {I}ntegration of
  {R}enewable {E}nergy {S}ources in the {G}erman {P}ower {S}upply {S}ystem from
  2015-2020 with an {O}utlook to 2025}.
\newblock \bibinfo{howpublished}{German Energy Agency. Final Report};
  \bibinfo{year}{2010}.
%Type = Misc
\bibitem[{{National Renewable Energy Laboratory}(2012)}]{NREL_RE_Futures}
\bibinfo{author}{{National Renewable Energy Laboratory}\xfnm[]}.
\newblock \bibinfo{title}{Renewable energy futures study}.
\newblock \bibinfo{howpublished}{Hand, M.M.; Baldwin, S.; DeMeo, E.; Reilly,
  J.M.; Mai, T.; Arent, D.; Porro, G.; Meshek, M.; Sandor, D. eds. 4 vols.
  NREL/TP-6A20-52409. Golden, CO: National Renewable Energy Laboratory};
  \bibinfo{year}{2012}.
\newblock \bibinfo{note}{[Online]:
  \url{http://www.nrel.gov/analysis/re_futures/}. Last accessed: 20/6/2012}.
%Type = Misc
\bibitem[{Czisch(2006)}]{Czisch_eurosupergrid}
\bibinfo{author}{Czisch\xfnm[ G.]}.
\newblock \bibinfo{title}{Low cost but totally renewable electricity supply for
  a huge supply area | a {E}uropean/{T}rans-{E}uropean example}.
\newblock \bibinfo{howpublished}{Unpublished manuscript}; \bibinfo{year}{2006}.
\newblock \bibinfo{note}{[Online]:
  \url{http://www.iset.uni-kassel.de/abt/w3-w/projekte/LowCostEuropElSup_revised_for_AKE_2006.pdf}.
  Last accessed: 20/6/2012}.
%Type = Book
\bibitem[{Czisch(2011)}]{Czisch_Book_IET}
\bibinfo{author}{Czisch\xfnm[ G.]}.
\newblock \bibinfo{title}{Scenarios for a Future Electricity Supply:
  cost-optimized variations on supplying {E}urope and its neighbours with
  electricity from renewable energies}.
\newblock \bibinfo{publisher}{Institution of Engineering and Technology
  {(IET)}}; \bibinfo{year}{2011}.
%Type = Misc
\bibitem[{{WWF}(2011)}]{WWF_energyreport}
\bibinfo{author}{{WWF}\xfnm[]}.
\newblock \bibinfo{title}{The energy report -- 100\% renewable energy by 2050}.
\newblock \bibinfo{howpublished}{World Wide Fund for Nature International and
  Ecofys}; \bibinfo{year}{2011}.
\newblock \bibinfo{note}{[Online]:
  \url{http://wwf.panda.org/what_we_do/footprint/climate_carbon_energy/energy_solutions/renewable_energy/sustainable_energy_report/}.
  Last accessed: 20/6/2012}.
%Type = Article
\bibitem[{Jacobson and Delucchi(2011)}]{WWSglobal_generation}
\bibinfo{author}{Jacobson\xfnm[ M.Z.]}, \bibinfo{author}{Delucchi\xfnm[ M.A.]}.
\newblock \bibinfo{title}{Providing all global energy with wind, water, and
  solar power, {P}art {I}: Technologies, energy resources, quantities and areas
  of infrastructure, and materials}.
\newblock \bibinfo{journal}{Energy Policy}
  \bibinfo{year}{2011};\bibinfo{volume}{39}(\bibinfo{number}{3}):\bibinfo{pages}{1154
  -- 1169}.
%Type = Article
\bibitem[{Archer and Jacobson(2007)}]{archer_jacobson}
\bibinfo{author}{Archer\xfnm[ C.L.]}, \bibinfo{author}{Jacobson\xfnm[ M.Z.]}.
\newblock \bibinfo{title}{Supplying baseload power and reducing transmission
  requirements by interconnecting wind farms}.
\newblock \bibinfo{journal}{Journal of Applied Meteorology and Climatology}
  \bibinfo{year}{2007};\bibinfo{volume}{46}:\bibinfo{pages}{1701 -- 1717}.
%Type = Inproceedings
\bibitem[{Paris et~al.(1984)Paris, Zini, Valtorta, Mazoni, Invernizzi,
  {DeFranco} et~al.}]{paris_longtrans}
\bibinfo{author}{Paris\xfnm[ L.]}, \bibinfo{author}{Zini\xfnm[ G.]},
  \bibinfo{author}{Valtorta\xfnm[ M.]}, \bibinfo{author}{Mazoni\xfnm[ G.]},
  \bibinfo{author}{Invernizzi\xfnm[ A.]}, \bibinfo{author}{{DeFranco}\xfnm[
  N.]}, et~al.
\newblock \bibinfo{title}{Present {L}imits of {V}ery {L}ong {T}ransmission
  {S}ystems}.
\newblock In: \bibinfo{booktitle}{International Conference on Large High
  Voltage Electric Systems (Cigre)}. \bibinfo{year}{1984},.
%Type = Article
\bibitem[{Paris(1992)}]{paris_grandinga}
\bibinfo{author}{Paris\xfnm[ L.]}.
\newblock \bibinfo{title}{Grand {I}nga {C}ase}.
\newblock \bibinfo{journal}{Power Engineering Review, IEEE}
  \bibinfo{year}{1992};\bibinfo{volume}{12}(\bibinfo{number}{6}):\bibinfo{pages}{13
  -- 17}.
%Type = Article
\bibitem[{Hammons et~al.(1993)Hammons, Olsen, Kacejko and Leung}]{Iceland_UK}
\bibinfo{author}{Hammons\xfnm[ T.]}, \bibinfo{author}{Olsen\xfnm[ A.]},
  \bibinfo{author}{Kacejko\xfnm[ P.]}, \bibinfo{author}{Leung\xfnm[ C.]}.
\newblock \bibinfo{title}{Proposed {I}celand/{U}nited {K}ingdom power link |
  {A}n indepth analysis of issues and returns}.
\newblock \bibinfo{journal}{IEEE Transactions on Energy Conversion}
  \bibinfo{year}{1993};\bibinfo{volume}{8}(\bibinfo{number}{3}):\bibinfo{pages}{566
  -- 575}.
%Type = Misc
\bibitem[{{The Guardian}(2012)}]{iceland_uk_guardian}
\bibinfo{author}{{The Guardian}\xfnm[]}.
\newblock \bibinfo{title}{Iceland's volcanoes may power {UK}}.
\newblock \bibinfo{year}{2012}.
\newblock \bibinfo{note}{[Online]:
  \url{http://www.guardian.co.uk/environment/2012/apr/11/iceland-volcano-green-power}.
  Last accessed: 20/6/2012}.
%Type = Article
\bibitem[{Boute and Willems(2012)}]{Rustec}
\bibinfo{author}{Boute\xfnm[ A.]}, \bibinfo{author}{Willems\xfnm[ P.]}.
\newblock \bibinfo{title}{{RUSTEC}: Greening {E}urope's energy supply by
  developing {R}ussia's renewable energy potential}.
\newblock \bibinfo{journal}{Energy Policy}
  \bibinfo{year}{2012};\bibinfo{volume}{51}(\bibinfo{number}{0}):\bibinfo{pages}{618
  -- 629}.
%Type = Book
\bibitem[{IEA(2010)}]{weo2010}
\bibinfo{author}{IEA\xfnm[]}.
\newblock \bibinfo{title}{World Energy Outlook}.
\newblock \bibinfo{publisher}{International Energy Agency};
  \bibinfo{year}{2010}.
%Type = Article
\bibitem[{Rudenko and Yershevich(1991)}]{intercont_grid_russia}
\bibinfo{author}{Rudenko\xfnm[ Y.]}, \bibinfo{author}{Yershevich\xfnm[ V.]}.
\newblock \bibinfo{title}{Is it possible and expedient to create a global
  energy network?}
\newblock \bibinfo{journal}{International Journal of Global Energy Issues}
  \bibinfo{year}{1991};\bibinfo{volume}{3}(\bibinfo{number}{3}).
%Type = Article
\bibitem[{Hammons et~al.(1994)Hammons, Falcon and
  Meisen}]{intercont_grid_hammons}
\bibinfo{author}{Hammons\xfnm[ T.J.]}, \bibinfo{author}{Falcon\xfnm[ J.A.]},
  \bibinfo{author}{Meisen\xfnm[ P.]}.
\newblock \bibinfo{title}{Remote renewable energy resources made possible by
  international electrical interconnections | {A} priority for all continents}.
\newblock \bibinfo{journal}{Power Generation Technology} \bibinfo{year}{1994};.
%Type = Article
\bibitem[{Kuwano(1994)}]{Kuwano_genesis}
\bibinfo{author}{Kuwano\xfnm[ Y.]}.
\newblock \bibinfo{title}{The {PV} era is coming | {T}he way to {GENESIS}}.
\newblock \bibinfo{journal}{Solar Energy Materials and Solar Cells}
  \bibinfo{year}{1994};\bibinfo{volume}{34}(\bibinfo{number}{1-4}):\bibinfo{pages}{27
  -- 39}.
%Type = Inproceedings
\bibitem[{Aboumahboub et~al.(2010)Aboumahboub, Schaber, Tzscheutschler and
  Hamacher}]{aboumahboub_Global_EU_grid}
\bibinfo{author}{Aboumahboub\xfnm[ T.]}, \bibinfo{author}{Schaber\xfnm[ K.]},
  \bibinfo{author}{Tzscheutschler\xfnm[ P.]}, \bibinfo{author}{Hamacher\xfnm[
  T.]}.
\newblock \bibinfo{title}{Optimization of the utilization of renewable energy
  sources in the electricity sector}.
\newblock In: \bibinfo{booktitle}{Proceedings of the 5th IASME / WSEAS
  International Conference on Energy \& Environment (EE '10)}.
  \bibinfo{year}{2010},.
%Type = Misc
\bibitem[{Wikipedia(2012)}]{airborne_wind}
\bibinfo{author}{Wikipedia\xfnm[]}.
\newblock \bibinfo{title}{Airborne wind turbine}.
\newblock \bibinfo{year}{ca. 2012}.
\newblock \bibinfo{note}{[Online]:
  \url{http://en.wikipedia.org/wiki/Airborne_wind_turbine}. Last accessed:
  20/6/2012}.
%Type = Misc
\bibitem[{{Statoil}(2011)}]{hywind}
\bibinfo{author}{{Statoil}\xfnm[]}.
\newblock \bibinfo{title}{Hywind | the world's first full-scale floating wind
  turbine}.
\newblock \bibinfo{year}{2011}.
\newblock \bibinfo{note}{[Online]:
  \url{http://www.statoil.com/en/TechnologyInnovation/NewEnergy/RenewablePowerProduction/Offshore/Hywind/Pages/HywindPuttingWindPowerToTheTest.aspx}.
  Last accessed: 20/6/2012}.
%Type = Article
\bibitem[{Bolonkin and Cathcart(2008)}]{Cathcart_Antarctica}
\bibinfo{author}{Bolonkin\xfnm[ A.]}, \bibinfo{author}{Cathcart\xfnm[ R.]}.
\newblock \bibinfo{title}{Antarctica: a southern hemisphere wind power
  station?}
\newblock \bibinfo{journal}{International Journal of Global Environmental
  Issues}
  \bibinfo{year}{2008};\bibinfo{volume}{8}(\bibinfo{number}{3}):\bibinfo{pages}{262
  -- 273}.
%Type = Inproceedings
\bibitem[{Sch\"{o}ffner et~al.(2006)Sch\"{o}ffner, Kunze and
  Smith}]{GIL_maxlength}
\bibinfo{author}{Sch\"{o}ffner\xfnm[ G.]}, \bibinfo{author}{Kunze\xfnm[ D.]},
  \bibinfo{author}{Smith\xfnm[ I.]}.
\newblock \bibinfo{title}{Gas insulated transmission lines | {S}uccessful
  underground bulk power transmission for more than 30 years}.
\newblock In: \bibinfo{booktitle}{{T}he 8th {IEE} {I}nternational {C}onference
  on {AC} and {DC} Power Transmission}. \bibinfo{year}{2006}, p.
  \bibinfo{pages}{271 -- 275}.
%Type = Misc
\bibitem[{UCTE(2008)}]{ucte_europe_russia}
\bibinfo{author}{UCTE\xfnm[]}.
\newblock \bibinfo{title}{Feasibility study: Synchronous interconnection of the
  {IPS/UPS} with the {UCTE}}.
\newblock \bibinfo{howpublished}{Summary of Investigations and Conclusions.
  Union for the Coordination of Transmission of Electricity. Brussels,
  Belgium.}; \bibinfo{year}{2008}.
%Type = Misc
\bibitem[{{E.ON UK}(2011)}]{eon_uk}
\bibinfo{author}{{E.ON UK}\xfnm[]}.
\newblock \bibinfo{title}{A {E}uropean {S}upergrid}.
\newblock \bibinfo{howpublished}{Memorandum {S}ubmitted by E.ON UK (ESG 05) to
  the {UK} {P}arliament}; \bibinfo{year}{2011}.
\newblock \bibinfo{note}{[Online]: \url{
  http://www.publications.parliament.uk/pa/cm201012/cmselect/cmenergy/writev/1040/esg05.htm}.
  Last accessed: 20/6/2012}.
%Type = Misc
\bibitem[{{Government of Greenland}(2009)}]{greenland_hydro}
\bibinfo{author}{{Government of Greenland}\xfnm[]}.
\newblock \bibinfo{title}{Invest in {G}reenland}.
\newblock \bibinfo{year}{2009}.
\newblock \bibinfo{note}{[Online]:
  \url{http://uk.nanoq.gl/Emner/Government/~/media/018C30DDD63E44B1BDC3A612E8FACD0F.ashx}.
  Last accessed: 20/6/2012}.
%Type = Misc
\bibitem[{{U.S. Energy Information Administration (EIA)}(2012)}]{eia_greenland}
\bibinfo{author}{{U.S. Energy Information Administration (EIA)}\xfnm[]}.
\newblock \bibinfo{title}{Greenland energy statistics}.
\newblock \bibinfo{year}{ca. 2012}.
\newblock \bibinfo{note}{[Online]:
  \url{http://www.eia.gov/countries/country-data.cfm?fips=GL}. Last accessed:
  20/6/2012}.
%Type = Article
\bibitem[{Delucchi and Jacobson(2011)}]{WWSglobal_grid}
\bibinfo{author}{Delucchi\xfnm[ M.A.]}, \bibinfo{author}{Jacobson\xfnm[ M.Z.]}.
\newblock \bibinfo{title}{Providing all global energy with wind, water, and
  solar power, {P}art {II}: Reliability, system and transmission costs, and
  policies}.
\newblock \bibinfo{journal}{Energy Policy}
  \bibinfo{year}{2011};\bibinfo{volume}{39}(\bibinfo{number}{3}):\bibinfo{pages}{1170
  -- 1190}.
%Type = Misc
\bibitem[{{Claverton Energy}(2010)}]{claverton_greenland}
\bibinfo{author}{{Claverton Energy}\xfnm[]}.
\newblock \bibinfo{title}{Prospects for trans-atlantic undersea power
  transmission}.
\newblock \bibinfo{year}{ca. 2010}.
\newblock \bibinfo{note}{[Online]:
  \url{http://www.claverton-energy.com/prospects-for-trans-atlantic-undersea-power-transmission.html}.
  Last accessed: 20/6/2012}.
%Type = Inproceedings
\bibitem[{Milligan et~al.(2010)Milligan, Donohoo, Lew, Ela, Kirby, Holttinen
  et~al.}]{reserves_comp}
\bibinfo{author}{Milligan\xfnm[ M.]}, \bibinfo{author}{Donohoo\xfnm[ P.]},
  \bibinfo{author}{Lew\xfnm[ D.]}, \bibinfo{author}{Ela\xfnm[ E.]},
  \bibinfo{author}{Kirby\xfnm[ B.]}, \bibinfo{author}{Holttinen\xfnm[ H.]},
  et~al.
\newblock \bibinfo{title}{Operating {R}eserves and {W}ind {P}ower
  {I}ntegration: {A}n {I}nternational {C}omparison}.
\newblock In: \bibinfo{booktitle}{The 9th {A}nnual {I}nternational {W}orkshop
  on {L}arge-{S}cale {I}ntegration of {W}ind {P}ower into {P}ower {S}ystems as
  well as on {T}ransmission {N}etworks for {O}ffshore {W}ind {P}ower {P}lants
  {C}onference}. \bibinfo{year}{2010},.
%Type = Misc
\bibitem[{ENTSO-E(2009)}]{ENTSOE_OpHandbook}
\bibinfo{author}{ENTSO-E\xfnm[]}.
\newblock \bibinfo{title}{Operation handbook, policy 3: Operational security}.
\newblock \bibinfo{howpublished}{European Network of Transmission System
  Operators for Electricity}; \bibinfo{year}{2009}.
%Type = Misc
\bibitem[{{Siemens}(2011)}]{HVDC_efficiency_siemens}
\bibinfo{author}{{Siemens}\xfnm[]}.
\newblock \bibinfo{title}{Ultra {HVDC} transmission system}.
\newblock \bibinfo{year}{ca. 2011}.
\newblock \bibinfo{note}{[Online]:
  \url{http://www.energy.siemens.com/hq/en/power-transmission/hvdc/hvdc-ultra/#content=Benefits}.
  Last accessed: 20/6/2012}.
%Type = Misc
\bibitem[{Nicolosi(2010)}]{nicolosi_negative_prices}
\bibinfo{author}{Nicolosi\xfnm[ M.]}.
\newblock \bibinfo{title}{Wind power integration, negative prices and power
  system flexibility | an empirical analysis of extreme events in {G}ermany}.
\newblock \bibinfo{howpublished}{EWI Working Paper, No. 10/01. Institute of
  Energy Economics at the University of Cologne. Germany.};
  \bibinfo{year}{2010}.
%Type = Article
\bibitem[{Parail(2009)}]{NorNed_effectOnPrice}
\bibinfo{author}{Parail\xfnm[ V.]}.
\newblock \bibinfo{title}{Can merchant interconnectors deliver lower and more
  stable prices? {T}he case of {N}or{N}ed}.
\newblock \bibinfo{journal}{EPRG Working Paper, Cambridge Working Paper on
  Economics, Faculty of Economics University of Cambridge}
  \bibinfo{year}{2009};.
%Type = Article
\bibitem[{Elliott(2012)}]{Elliott2012}
\bibinfo{author}{Elliott\xfnm[ D.]}.
\newblock \bibinfo{title}{Emergence of {E}uropean supergrids -- {E}ssay on
  strategy issues}.
\newblock \bibinfo{journal}{Energy Strategy Reviews}
  \bibinfo{year}{2012};(\bibinfo{number}{0}):\bibinfo{pages}{--}.
\newblock
  \urlprefix\url{http://www.sciencedirect.com/science/article/pii/S2211467X12000120}.
%Type = Misc
\bibitem[{OffshoreGrid(2011)}]{Offshore_report}
\bibinfo{author}{OffshoreGrid\xfnm[]}.
\newblock \bibinfo{title}{Offshore electricity grid infrastructure in
  {E}urope}.
\newblock \bibinfo{howpublished}{A Techno-Economic Assessment, Final Report};
  \bibinfo{year}{2011}.
%Type = Misc
\bibitem[{{w}nn(2010)}]{wnn_olkiluoto}
\bibinfo{author}{{w}nn\xfnm[]}.
\newblock \bibinfo{title}{Start-up of {F}innish {EPR} pushed back to 2013}.
\newblock \bibinfo{howpublished}{World {N}uclear {N}ews}; \bibinfo{year}{2010}.
\newblock \bibinfo{note}{[Online]:
  \url{http://www.world-nuclear-news.org/NN-Startup_of_Finnish_EPR_pushed_back_to_2013-0806104.html}.
  Last accessed: 20/6/2012}.
%Type = Misc
\bibitem[{Bogan(2008)}]{forbes_perdido}
\bibinfo{author}{Bogan\xfnm[ J.]}.
\newblock \bibinfo{title}{Shell's {R}adical {R}ig}.
\newblock \bibinfo{howpublished}{Forbes Magazine}; \bibinfo{year}{2008}.
\newblock \bibinfo{note}{[Online]:
  \url{http://www.forbes.com/forbes/2008/1124/072.html}. Last accessed:
  20/6/2012}.
%Type = Misc
\bibitem[{{EC}(2011)}]{EC_Invest_Needs}
\bibinfo{author}{{EC}\xfnm[]}.
\newblock \bibinfo{title}{Energy infrastructure investment needs and financing
  requirements}.
\newblock \bibinfo{howpublished}{SEC(2011)755, European Commission};
  \bibinfo{year}{2011}.
%Type = Misc
\bibitem[{{EC}(2006)}]{ECgreenpaper}
\bibinfo{author}{{EC}\xfnm[]}.
\newblock \bibinfo{title}{A {E}uropean strategy for sustainable, competitive
  and secure energy}.
\newblock \bibinfo{howpublished}{Green Paper. {C}ommission of the {E}uropean
  {C}ommunities.}; \bibinfo{year}{2006}.
%Type = Misc
\bibitem[{Barry and Wheelock(2010)}]{pikeresearch_investments}
\bibinfo{author}{Barry\xfnm[ D.]}, \bibinfo{author}{Wheelock\xfnm[ C.]}.
\newblock \bibinfo{title}{Electricity transmission infrastructure | market
  drivers and barriers, emerging technologies, key industry players, and
  worldwide growth forecasts}.
\newblock \bibinfo{howpublished}{Research report, Pike Research. Executive
  Summary}; \bibinfo{year}{2010}.
%Type = Misc
\bibitem[{{Statnett}(2008)}]{norned_press}
\bibinfo{author}{{Statnett}\xfnm[]}.
\newblock \bibinfo{title}{Nor{N}ed cable off to a promising start}.
\newblock \bibinfo{howpublished}{Press release}; \bibinfo{year}{2008}.
\newblock \bibinfo{note}{[Online]:
  \url{http://www.statnett.no/en/News/News-archive-Temp/News-archive-2008/NorNed-cable-off-to-a-promising-start-/}.
  Last accessed: 20/6/2012}.
%Type = Misc
\bibitem[{Chatzivasileiadis(2012)}]{SCH_techreport}
\bibinfo{author}{Chatzivasileiadis\xfnm[ S.]}.
\newblock \bibinfo{title}{Transmission investments in deregulated electricity
  markets}.
\newblock \bibinfo{howpublished}{Technical Report, ETH Zurich, EEH Power
  Systems Laboratory}; \bibinfo{year}{2012}.
\newblock \bibinfo{note}{[Online]:
  \url{http://e-collection.library.ethz.ch/view/eth:5559}. Last accessed:
  20/6/2012}.
%Type = Article
\bibitem[{Joskow(2006)}]{Joskow_incent_long}
\bibinfo{author}{Joskow\xfnm[ P.]}.
\newblock \bibinfo{title}{Incentive regulation in theory and practice:
  Electricity distribution and transmission networks}.
\newblock \bibinfo{journal}{Prepared for National Bureau of Economic Research
  Economic Regulation Project} \bibinfo{year}{2006};.
%Type = Misc
\bibitem[{Hogan(2003)}]{hogan}
\bibinfo{author}{Hogan\xfnm[ W.W.]}.
\newblock \bibinfo{title}{Transmission market design}.
\newblock \bibinfo{howpublished}{KSG Working Paper No. RWP03-040};
  \bibinfo{year}{2003}.
\newblock \bibinfo{note}{[Online]: \url{http://ssrn.com/abstract=453483}. Last
  accessed: 20/6/2012}.
%Type = Article
\bibitem[{Joskow and Tirole(2005)}]{Joskow_Tirole_MTI}
\bibinfo{author}{Joskow\xfnm[ P.]}, \bibinfo{author}{Tirole\xfnm[ J.]}.
\newblock \bibinfo{title}{Merchant transmission investment}.
\newblock \bibinfo{journal}{The Journal of Industrial Economics}
  \bibinfo{year}{2005};\bibinfo{volume}{53}(\bibinfo{number}{2}):\bibinfo{pages}{233
  -- 264}.
%Type = Book
\bibitem[{EC(2003)}]{EC_2003}
\bibinfo{author}{EC\xfnm[]}.
\newblock \bibinfo{title}{Regulations on conditions for access to the network
  for cross-border exchanges in electricity}.
\newblock \bibinfo{publisher}{European Commission, Brussels};
  \bibinfo{year}{2003}.
%Type = Misc
\bibitem[{Mullett(2010)}]{NordBalt_totalcosts}
\bibinfo{author}{Mullett\xfnm[ A.]}.
\newblock \bibinfo{title}{Brussels allocates {EUR} 131 mln for {N}ord{B}alt}.
\newblock \bibinfo{howpublished}{Press Release}; \bibinfo{year}{2010}.
\newblock \bibinfo{note}{[Online]: \url{http://balticreports.com/?p=22711}.
  Last accessed: 20/6/2012}.
%Type = Article
\bibitem[{Brunekreeft(2004)}]{Brunekreeft_Cambridge}
\bibinfo{author}{Brunekreeft\xfnm[ G.]}.
\newblock \bibinfo{title}{Market-based investment in electricity transmission
  networks: {C}ontrollable flow}.
\newblock \bibinfo{journal}{Utilities Policy}
  \bibinfo{year}{2004};\bibinfo{volume}{12}(\bibinfo{number}{4}):\bibinfo{pages}{269
  -- 281}.
%Type = Misc
\bibitem[{J\"{a}derberg(2010{\natexlab{a}})}]{NordBalt_cable}
\bibinfo{author}{J\"{a}derberg\xfnm[ M.]}.
\newblock \bibinfo{title}{Svenska {K}raftn\"{a}t: {P}rocurement of
  {N}ord{B}alt's cable supply and installation complete}.
\newblock \bibinfo{howpublished}{Press Release};
  \bibinfo{year}{2010}{\natexlab{a}}.
\newblock \bibinfo{note}{[Online]: \url{http://www.svk.se/Start/English/News/}.
  Last accessed: 20/6/2012}.
%Type = Phdthesis
\bibitem[{Krause(2007)}]{krause_phd}
\bibinfo{author}{Krause\xfnm[ T.]}.
\newblock \bibinfo{title}{Evaluating congestion management schemes in
  liberalized electricity markets applying agent-based computational
  economics}.
\newblock Ph.D. thesis; ETH Zurich, Switzerland; \bibinfo{year}{2007}.
%Type = Phdthesis
\bibitem[{Kurzidem(2010)}]{kurzidem_phd}
\bibinfo{author}{Kurzidem\xfnm[ M.J.]}.
\newblock \bibinfo{title}{Analysis of {F}low-based {M}arket {C}oupling in
  {O}ligopolistic {P}ower {M}arkets}.
\newblock Ph.D. thesis; ETH Zurich, Switzerland; \bibinfo{year}{2010}.
%Type = Article
\bibitem[{Dinica(2011)}]{Dinica}
\bibinfo{author}{Dinica\xfnm[ V.]}.
\newblock \bibinfo{title}{Renewable electricity production costs | a framework
  to assist policy-makers' decisions on price support}.
\newblock \bibinfo{journal}{Energy Policy}
  \bibinfo{year}{2011};\bibinfo{volume}{39}(\bibinfo{number}{7}):\bibinfo{pages}{4153
  -- 4167}.
%Type = Misc
\bibitem[{{Pipelines International}(2012)}]{centralasia_china}
\bibinfo{author}{{Pipelines International}\xfnm[]}.
\newblock \bibinfo{title}{Construction on third line begins for central
  asia-china gas pipeline}.
\newblock \bibinfo{year}{2012}.
\newblock \bibinfo{note}{[Online]:
  \url{http://pipelinesinternational.com/news/construction_on_third_line_begins_for_central_asia-china_gas_pipeline/066998/}.
  Last accessed: 20/6/2012}.
%Type = Misc
\bibitem[{Hauth et~al.(1997)Hauth, Tatro, Railing, Johnson, Stewart and
  Fink}]{Hauth_HVDCcosts}
\bibinfo{author}{Hauth\xfnm[ R.L.]}, \bibinfo{author}{Tatro\xfnm[ P.J.]},
  \bibinfo{author}{Railing\xfnm[ B.D.]}, \bibinfo{author}{Johnson\xfnm[ B.K.]},
  \bibinfo{author}{Stewart\xfnm[ J.R.]}, \bibinfo{author}{Fink\xfnm[ J.L.]}.
\newblock \bibinfo{title}{{HVDC} power transmission technology assessment}.
\newblock \bibinfo{howpublished}{ORNLSub/95-SR893/1, Oak Ridge National
  Laboratory, Oak Ridge, TN}; \bibinfo{year}{1997}.
\newblock \bibinfo{note}{[Online]:
  \url{www.osti.gov/bridge/product.biblio.jsp?osti_id=580574}. Last accessed:
  20/6/2012}.
%Type = Article
\bibitem[{Weigt et~al.(2010)Weigt, Jeske, Leuthold and von
  Hirschhausen}]{Weigt_eurosupergrid}
\bibinfo{author}{Weigt\xfnm[ H.]}, \bibinfo{author}{Jeske\xfnm[ T.]},
  \bibinfo{author}{Leuthold\xfnm[ F.]}, \bibinfo{author}{von Hirschhausen\xfnm[
  C.]}.
\newblock \bibinfo{title}{``{T}ake the long way down'': Integration of
  large-scale {N}orth {S}ea wind using {HVDC} transmission}.
\newblock \bibinfo{journal}{Energy Policy}
  \bibinfo{year}{2010};\bibinfo{volume}{38}(\bibinfo{number}{7}):\bibinfo{pages}{3164
  -- 3173}.
%Type = Misc
\bibitem[{{DLR}(2006)}]{DLR_HVDCcosts}
\bibinfo{author}{{DLR}\xfnm[]}.
\newblock \bibinfo{title}{Trans-{M}editerranean interconnection for
  concentrating solar power}.
\newblock \bibinfo{howpublished}{German Aerospace Center, Institute of
  Technical Thermodynamics, Section Systems Analysis and Technology Assessment.
  Study commisioned by the {F}ederal {M}inistry for the {E}nvironment, {N}ature
  {C}onservation and {N}uclear {S}afety. Germany}; \bibinfo{year}{2006}.
%Type = Misc
\bibitem[{{NorNed}(2008{\natexlab{a}})}]{NorNed_costs}
\bibinfo{author}{{NorNed}\xfnm[]}.
\newblock \bibinfo{title}{The longest electricity cable in the world is
  operational}.
\newblock \bibinfo{howpublished}{Press Release};
  \bibinfo{year}{2008}{\natexlab{a}}.
\newblock \bibinfo{note}{[Online]:
  \url{http://www.norned-auction.org/news/newsitems/The_longest_electricity_cable_in_the_world_is_operational.aspx}.
  Last accessed: 20/6/2012}.
%Type = Misc
\bibitem[{{TenneT}(2011)}]{BritNed_costs}
\bibinfo{author}{{TenneT}\xfnm[]}.
\newblock \bibinfo{title}{Brit{N}ed cable electrically connects {U}nited
  {K}ingdom and the {N}etherlands}.
\newblock \bibinfo{howpublished}{Press Release}; \bibinfo{year}{2011}.
\newblock \bibinfo{note}{[Online]:
  \url{http://www.tennet.org/english/tennet/news/110401KabeltussenVerenigdKoninkrijksuccesvolinbedrijf.aspx}.
  Last accessed: 20/6/2012}.
%Type = Misc
\bibitem[{Wikipedia(2011)}]{NorGer_wiki}
\bibinfo{author}{Wikipedia\xfnm[]}.
\newblock \bibinfo{title}{Nor{G}er}.
\newblock \bibinfo{year}{ca. 2011}.
\newblock \bibinfo{note}{[Online]: \url{http://de.wikipedia.org/wiki/NorGer}.
  Last accessed: 20/6/2012}.
%Type = Misc
\bibitem[{J\"{a}derberg(2010{\natexlab{b}})}]{NordBalt_converters}
\bibinfo{author}{J\"{a}derberg\xfnm[ M.]}.
\newblock \bibinfo{title}{Svenska {K}raftn\"{a}t: {P}rocurement of
  {N}ord{B}alt's converter stations complete}.
\newblock \bibinfo{howpublished}{Press Release};
  \bibinfo{year}{2010}{\natexlab{b}}.
\newblock \bibinfo{note}{[Online]: \url{http://www.svk.se/Start/English/News/}.
  Last accessed: 20/6/2012}.
%Type = Article
\bibitem[{Negra et~al.(2006)Negra, Todorovic and
  Ackermann}]{Negra_HVDCconverter_efficiency}
\bibinfo{author}{Negra\xfnm[ N.B.]}, \bibinfo{author}{Todorovic\xfnm[ J.]},
  \bibinfo{author}{Ackermann\xfnm[ T.]}.
\newblock \bibinfo{title}{Loss evaluation of {HVAC} and {HVDC} transmission
  solutions for large offshore wind farms}.
\newblock \bibinfo{journal}{Electric Power Systems Research}
  \bibinfo{year}{2006};\bibinfo{volume}{76}(\bibinfo{number}{11}):\bibinfo{pages}{916
  -- 927}.
%Type = Misc
\bibitem[{{ABB}(2005)}]{ABB_HVDCavailability}
\bibinfo{author}{{ABB}\xfnm[]}.
\newblock \bibinfo{title}{{HVDC} {C}able {T}ransmissions}.
\newblock \bibinfo{howpublished}{ABB High Voltage Cables. ABB Power Systems};
  \bibinfo{year}{ca. 2005}.
%Type = Misc
\bibitem[{Skog(2004)}]{HVDC_lifetime}
\bibinfo{author}{Skog\xfnm[ J.E.]}.
\newblock \bibinfo{title}{{HVDC} {T}ransmission and {L}ife {E}xpectancy}.
\newblock \bibinfo{howpublished}{Memo Statnett-TenneT}; \bibinfo{year}{2004}.
\newblock \bibinfo{note}{[Online]:
  \url{http://www.tennet.org/english/images/19-UK-B7-HVDC_Transmission_and_Lifetime_Expectancy_tcm43-12302.pdf}.
  Last accessed: 20/6/2012}.
%Type = Misc
\bibitem[{Short et~al.(1995)Short, Packey and Holt}]{NREL_discountrate}
\bibinfo{author}{Short\xfnm[ W.]}, \bibinfo{author}{Packey\xfnm[ D.J.]},
  \bibinfo{author}{Holt\xfnm[ T.]}.
\newblock \bibinfo{title}{A manual for the economic evaluation of energy
  efficiency and renewable energy technologies}.
\newblock \bibinfo{howpublished}{National Renewable Energy Laboratory};
  \bibinfo{year}{1995}.
\newblock \bibinfo{note}{NREL/TP-462-5173}.
%Type = Misc
\bibitem[{{NorNed}(2008{\natexlab{b}})}]{Norned_auctionrules}
\bibinfo{author}{{NorNed}\xfnm[]}.
\newblock \bibinfo{title}{Rules for the auctioning of physical electricity
  transfer rights on the {HVDC}-link between the {N}etherlands and {N}orway}.
\newblock \bibinfo{year}{2008}{\natexlab{b}}.
\newblock \bibinfo{note}{[Online]:
  \url{http://www.norned-auction.org/rulesregulations/default.aspx}. Last
  accessed: 20/6/2012}.
%Type = Inproceedings
\bibitem[{Kurokawa et~al.(2002)Kurokawa, Kato, Ito, Komoto, Kichimi and
  Sugihara}]{kurokawa_vlspv}
\bibinfo{author}{Kurokawa\xfnm[ K.]}, \bibinfo{author}{Kato\xfnm[ K.]},
  \bibinfo{author}{Ito\xfnm[ M.]}, \bibinfo{author}{Komoto\xfnm[ K.]},
  \bibinfo{author}{Kichimi\xfnm[ T.]}, \bibinfo{author}{Sugihara\xfnm[ H.]}.
\newblock \bibinfo{title}{A cost analysis of very large scale {PV (VLS-PV)}
  system on the world deserts}.
\newblock In: \bibinfo{booktitle}{Proceedings of the Twenty-Ninth IEEE
  Photovoltaic Specialists Conference}. \bibinfo{year}{2002}, p.
  \bibinfo{pages}{1672 -- 1675}.

\end{thebibliography}

\end{document}